\documentclass[12pt]{article}
\usepackage{tocloft}
\usepackage{amsfonts}
\usepackage{amsmath}
\usepackage{amssymb}
\usepackage{array,slashed}
\usepackage{bigints}
\usepackage{bm}
\usepackage{bbm}
\usepackage{upgreek}
\usepackage{booktabs}
\usepackage[nosort]{cite}
\usepackage{color}
\usepackage{dsfont}
\usepackage{float}
\usepackage{framed}
\usepackage{graphicx}
\usepackage{indentfirst}
\usepackage{mathrsfs}
\usepackage{multirow}
\usepackage{pdflscape}
\usepackage{setspace}
\usepackage{subdepth}
\usepackage{subfig}
\usepackage{titlesec}
\usepackage{wrapfig}
\usepackage[all]{xy}
\usepackage{young}
\usepackage[vcentermath]{youngtab}
\usepackage{relsize}
\usepackage{stackengine}
\usepackage{verbatim}
\usepackage{slashed}

\newcommand{\re}{{\rm e}}
\newcommand{\ri}{{\rm i}}
\newcommand{\rd}{{\rm d}}

\usepackage{hyperref}
\hypersetup{colorlinks=true}
\hypersetup{linkcolor=black}
\hypersetup{citecolor=black}
\hypersetup{urlcolor=black}

\numberwithin{equation}{section}

\usepackage[left=2.5cm,right=2.5cm,top=2.5cm,bottom=3cm]{geometry}
\fontdimen2\font=1.2\fontdimen2{\jot}{5pt}

\titleformat{\section}{\large\bfseries}{\thesection.}{4pt}{}
\titlespacing{\section}{0pt}{20pt}{6pt}

\titleformat{\subsection}{\normalfont\bfseries}{\thesubsection.}{4pt}{}
\titlespacing{\subsection}{0pt}{15pt}{6pt}

\titleformat{\subsubsection}{\normalfont\itshape}{\thesubsubsection.}{4pt}{}
\titlespacing{\subsubsection}{0pt}{15pt}{6pt}

\titleformat{\paragraph}{\normalfont\itshape}{\theparagraph.}{4pt}{}
\titlespacing{\paragraph}{0pt}{15pt}{6pt}

\def\bar{\overline}

\def\half{{1 \over 2}}

\def\1{{\mathds 1}}

\def\Im{\mathop{\rm Im}}

\DeclareMathOperator{\Tr}{\mathrm{Tr}}
\DeclareMathAlphabet{\mathbfsf}{OT1}{cmss}{bx}{n}

\newcommand{\R}{{\mathbb R}}
\newcommand{\ba}{\begin{aligned}}
\newcommand{\ea}{\end{aligned}}

\def\CN{{\mathcal N}}

\def\IR{{\mathbb R}}

\def\bR{\mathbb{R}}
\def\bC{\mathbb{C}}

\newcommand{\cF}{\mathcal{F}}

\newcommand{\cM}{\mathcal M}
\newcommand{\cN}{\mathcal{N}}
\newcommand{\cO}{\mathcal{O}}

\newcommand{\cR}{\mathcal{R}}

\def\osp{\mathfrak{osp}}
\def\su{\mathfrak{su}}
\def\sp{\mathfrak{sp}}
\def\u{\mathfrak{u}}

\def\O{\mathrm{O}}

\def\SU{\mathrm{SU}}

\newcommand\scrL{\mathscr{L}}
\newcommand{\ed}{\,.}
\newcommand{\ec}{\,,}

\newcommand{\be}{\begin{equation}}
\newcommand{\ee}{\end{equation}}

\newcommand{\inst}{\textrm{Inst}}

\DeclareFontShape{OT1}{cmr}{mx}{n}%
{<->cmr10}{}
\newcommand{\mytitlefont}{\fontseries{mx}\selectfont}
\DeclareMathAlphabet{\titlemath}{OT1}{cmr}{mx}{n}

\begin{document}
	
	
%

%
\begin{titlepage}
\begin{center}
~\\[2cm]
{\fontsize{29pt}{0pt} \mytitlefont Extremal Correlators and Random Matrix Theory}
~\\[1cm]
Alba Grassi, Zohar Komargodski, and Luigi Tizzano\hskip1pt
~\\[0.5cm]
{\it Simons Center for Geometry and Physics, SUNY, Stony Brook, NY 11794, USA}\\[0.2cm]
\end{center}
\vskip2cm
\noindent 

We study the correlation functions of Coulomb branch operators of four-dimensional $\mathcal{N}=2$ Superconformal Field Theories (SCFTs). We focus on rank-one theories, such as the $\SU(2)$ gauge theory with four fundamental hypermultiplets. ``Extremal" correlation functions, involving exactly one anti-chiral operator, are perhaps the simplest nontrivial correlation functions in four-dimensional Quantum Field Theory. We show that the large charge limit of extremal correlators is captured by a ``dual'' description which is a chiral random matrix model of the Wishart-Laguerre type. This gives an analytic handle on the physics in some particular excited states. In the limit of large random matrices we find the physics of a non-relativistic axion-dilaton effective theory. The random matrix model also admits a 't Hooft expansion in which the matrix is taken to be large and simultaneously the coupling is taken to zero. This explains why the extremal correlators of $\SU(2)$ gauge theory obey a nontrivial double scaling limit in states of large charge. We give an exact solution for the first two orders in the 't Hooft expansion of the random matrix model and compare with expectations from effective field theory, previous weak coupling results, and we analyze the non-perturbative terms in the strong 't Hooft coupling limit.  
Finally, we apply the random matrix theory techniques to study extremal correlators in rank-1 Argyres-Douglas theories. We compare our results with effective field theory and with some available numerical bootstrap bounds.

\vfill 
\begin{flushleft}
August 2019
\end{flushleft}
\end{titlepage}
%
		
	
\setcounter{tocdepth}{3}
\renewcommand{\cfttoctitlefont}{\large\bfseries}
\renewcommand{\cftsecaftersnum}{.}
\renewcommand{\cftsubsecaftersnum}{.}
\renewcommand{\cftsubsubsecaftersnum}{.}
\renewcommand{\cftdotsep}{6}
\renewcommand\contentsname{\centerline{Contents}}
	
\tableofcontents


\section{Introduction}

In this paper we revisit the properties of some of the simplest nontrivial correlation functions in four-dimensional quantum field theories. 
We will be discussing the ``extremal'' correlation functions in $\mathcal{N}=2$ supersymmetric conformal theories in four dimensions. These theories often have exactly marginal parameters, namely, there is a tunable coupling constant and the extremal correlation functions we will study depend on it. In the simplest examples, the tunable coupling constant is simply $g_{\rm YM}$, i.e. the Yang-Mills coupling (and the theta angle).

In general we can only hope to compute observables in perturbation theory in $g_{\rm YM}$ where the difficulty of the computation of Feynman diagrams increases rapidly with the order in perturbation theory. The extremal correlators have a nontrivial perturbative expansion but thanks to localization techniques the computation of higher-order terms is technically not as formidable as the computation of the usual Feynman diagrams.

The relative simplicity of extremal correlators allows us to explore various questions such as the resurgence properties of perturbation theory as well as the behavior of perturbation theory in the presence of large quantum numbers.  

The local operators which we consider are the Coulomb branch operators $\cO(x)$, which are annihilated by the $\bar Q$ supercharges (and are super-conformal primaries). Correlation functions of  $\cO$ at the conformal fixed point vanish due to global symmetry selection rules but as soon as we allow at least one insertion of $\bar\cO$ the correlation functions no longer vanish. 
If there is precisely one insertion of $\bar \cO$ the correlation function is called extremal. 

In the theories we focus on in this paper (mostly $\SU(2)$ gauge theory with four hypermultiplets in the fundamental representation) the set of Coulomb branch operators is labeled by non-negative integers $\cO_n$ such that the $\u(1)_R$ charge of $\cO_n$ is proportional to $n$. It is possible to normalize the OPE coefficients such that 
$$\cO_n\cO_m\sim \cO_{n+m}+{\rm regular}~.$$
The physical conformal field theory data is then in two-point correlation functions 
\begin{equation}\label{twoptin}\langle \cO_n(0)\bar\cO_n(\infty)\rangle = G_{2n}(g_{\rm YM},\theta)~.\end{equation}
Our goal is thus to compute the observables $G_{2n}(g_{\rm YM},\theta)$.

In several recent papers the main focus was on studying some properties of $G_{2n}(g_{\rm YM},\theta)$ at fixed $n$. The purpose of this was twofold. First, one could learn about the high-order behavior of perturbation theory, the interplay with instantons and Borel resummation. Second, the system of observables $G_{2n}$ obey a certain integrable structure which we will expand about in the main body of the paper. We will review all of these developments in the main body of the paper.

Here we shift our attention to a different class of problems which involve the large $n$ limit. This corresponds to the limit of large $\u(1)_R$ charge. 
In order to have a physical picture of this limit it is perhaps more convenient to normalize the two-point functions \eqref{twoptin} canonically. Then the three-point functions are given by 
$$\langle \cO_n(0)\cO_1(1)\bar\cO_{n+1}(\infty)\rangle = {1\over \sqrt G_2}\sqrt{G_{2n+2}\over G_{2n}}~.$$
In radial quantization we can think about this as a matrix element of a ``simple'' operator between a state of charge $\sim n$ and a state of charge $\sim n+1$. 
The operator $\cO_1$ is simple in the sense that its quantum numbers are fixed while $n$ may be large.

In the context of the Eigenstate Thermalization Hypothesis (ETH), it is widely expected that in nontrivial Quantum Field Theories correlation functions of ``simple'' operators in sufficiently generic states are captured by Random Matrix Theory. (This expectation is based mostly on simulations of many-body systems. In fact some of the earliest work on ETH already made contact with Random Matrix Theory. For some recent work see for instance~\cite{mondaini2017eigenstate,Dymarsky:2017zoc} and references therein.)  Here we are studying matrix elements of $\cO_1$ between states which are by no means generic.\footnote{If the radius of the sphere is $R$, the energy density of our states scales like $n/R^4$ and the charge density like $n/R^3$. Due to the Boltzmann factor $e^{{-H-\mu Q\over T}}$, at low temperatures only states with $H\leq |\mu Q|$ play a role. Typically for states with finite charge density the energy density is bounded from below and hence up to some $\mu^*$ the partition function at low temperatures is $\mu$ independent. This is not the case in our system, where there are states with finite charge density and arbitrarily small energy density.  In this sense these states are somewhat unusual. On the other hand, as we will see, there is much analytic control over these states, and, in some respects, they can be viewed as a playground for more realistic thermodynamic states.} However, we will see that a natural chiral Random Matrix Theory (RMT) description emerges! 
The ensemble of random matrices that we will encounter is the so-called chiral ensemble (knows as the Wishart-Laguerre or Altland-Zirnbauer ensemble \cite{Altland:1997zz}) of general complex matrices whose rank is related to the number of operator insertions. There is an emergent gauge symmetry which acts on the random matrices from the left and right independently. It becomes exactly Gaussian in the strict $n\to\infty$ limit. This ensemble also appears naturally in Quantum Chromodynamics~\cite{Shuryak:1992pi, Verbaarschot:1994qf, Verbaarschot:2000dy} and several other contexts such as~\cite{Douglas:2003up}.  

Since our system has adjustable coupling constants $g_{\rm YM},\theta$ one can actually take the large $n$ limit in several ways. 

A limit that always exists is fixed $g_{\rm YM},\theta$ and $n\to\infty$. In this limit the theory is expected to be described by a non-relativistic axion-dilaton effective Lagrangian~\cite{Hellerman:2017veg,Hellerman:2017sur,Hellerman:2018xpi}. (This is the supersymmetric version of the effective Lagrangian of~\cite{Hellerman:2015nra,Monin:2016jmo,Alvarez-Gaume:2016vff}, see also~\cite{Jafferis:2017zna} for a Bootstrap analysis.) In terms of the RMT  this limit corresponds to large matrices with a fixed potential. Remarkably, the two approaches agree!
The agreement is nontrivial. For instance, some loop effects in effective field theory are captured by edge effects in random matrix theory.

A more surprising and richer limit is 
\begin{equation}\label{newl}{1\over 4\pi}g_{\rm YM\,}^2n\equiv \lambda~,\quad n\to \infty\end{equation}
with fixed $\lambda$. It reminds of the 't Hooft limit in gauge theories, though here  we are studying just $\SU(2)$ gauge theory   and $n$ is the $\u(1)_R$ charge. It is nontrivial that this limit makes sense  from a gauge theory perspective.
As in the 't Hooft limit, gauge instantons are {\it naively} suppressed as $e^{-8\pi^2/ g_{\rm YM}^2} \rightarrow e^{-2\pi n/\lambda}$. We will see that the effective physics of this limit is not simply the axion-dilaton effective Lagrangian. The RMT approach immediately shows that \eqref{newl} makes sense since this is nothing but the familiar 't Hooft limit. Hence, in terms of RMT, $\lambda$ is the familiar 't Hooft parameter and the limit exists because the diagrams admit the usual genus expansion and double line notation. In this sense the RMT is really a dual description. Since the double scaling limit \eqref{newl} corresponds to a 't Hooft expansion in the RMT approach, we see that there is an underlying weakly coupled string world-sheet theory in the double scaling limit. We do not explore this aspect further in this paper.

An interesting limit is that of strong 't Hooft coupling $\lambda\to\infty$. 
We find in that limit an exponentially small piece $e^{-\sqrt{4\pi}\sqrt\lambda}$ which is reminiscent of a world-sheet instanton. In terms of the original gauge theory, this is due to a threshold effect of the hypermultiplet BPS particle. 

While in terms of the dual RMT the existence of the limit~\eqref{newl} is obvious, it is worth developing an intuitive picture for why this limit exists directly from the gauge theory point of view. Inserting $\cO_n$ in the gauge theory kicks the dilaton field away from the origin $\phi\sim \sqrt n$ (with the dimensionful scale being the radius of the sphere in radial quantization). The masses of the hypermultiplets and W-boson BPS states scale as $g_{\rm YM}\phi\sim g_{\rm YM}\sqrt n$.  The monopole masses on the other hand scale like ${1\over g_{\rm YM} } \phi \sim \sqrt n /g_{\rm YM}$. From these facts we see that at fixed $g_{\rm YM}$ and large $n$ all the BPS particles decouple and we are left with the axion-dilaton field with a nontrivial profile which breaks Lorentz symmetry. This is essentially the underlying reason for the non-relativistic axion-dilaton theory at large $n$ and fixed $g_{\rm YM}$.  However, we see that if we keep $g_{\rm YM}^2n$ fixed then the hypermultiplet and W-boson have a fixed nonzero mass while the magnetically charged particles decouple. Also any possible multi-particle states made out of the hypermultiplets and W-boson survive the limit of fixed $g_{\rm YM}^2n$. At small $\lambda$ the physics is rather complicated since the electric BPS particles are light. 
At large $\lambda$ it becomes increasingly improbable to excite them and instead they lead to contact terms along with exponentially small contributions, such as the $e^{-\sqrt{4\pi}\sqrt\lambda}$ that we mentioned. The RMT approach allows us to understand this physics in detail. 

These results raise various interesting questions that go beyond the realm of supersymmetric gauge theories. For instance, we can ask about the large charge limit of the $\O(2)$ Wilson-Fisher fixed point. An insertion of $\phi^n$ (where $\phi$ is the complex scalar field in the $\O(2)$ Lagrangian) kicks the field to $\phi\sim \sqrt n$ as before. The mass of the radial excitation in that region is determined from the quartic coupling to be $\lambda (\sqrt n)^2\sim \lambda n\sim \epsilon n$, where in the last step we used that the quartic coupling scales like $\epsilon=4-d$. In analogy to what we find in this work, it is therefore tempting to explore the same kind of 't Hooft-like expansion in the $\O(2)$ Wilson Fisher fixed point. For related work see~\cite{Libanov:1994ug, Libanov:1995gh, Son:1995wz, Appear:2019}.\footnote{We thank M. Serone for a useful discussion on the subject.} The ordinary large charge limit is described by the non-relativistic Goldstone mode but the double scaling limit should be described by a more intricate theory that includes the radial mode. It would be really interesting to see if there is a dual matrix model that translates the double scaling limit to a standard 't Hooft limit.

In this paper we limit ourselves to a detailed analysis of the $\cN=2$ supersymmetric rank-1 theories.\footnote{Certain 2d and 3d supersymmetric theories also admit some similar classes of protected operators and extremal correlation functions can be defined~\cite{Dedushenko:2016jxl,Chen:2017fvl,Ishtiaque:2017trm,Dedushenko:2017avn}. It would be interesting to consider these examples along the lines of the discussion here. Likewise, there has been some work on the extremal correlators in higher-rank theories, e.g.,~\cite{Rodriguez-Gomez:2016ijh,Baggio:2016skg} and it would be very nice to understand the large charge limit in that case. Finally, there have been tantalizing hints that some of the structure carries over to non-conformal theories~\cite{Billo:2017glv,Billo:2019job}, and it would be nice to understand this better.}

The outline of the paper is as follows. In section 2 we review some of the literature on extremal correlators with the aim of making the presentation essentially self-contained. In the same section we review the predictions of effective field theory for the behavior of extremal correlators at large charge and fixed coupling.
In section 3 we develop the Random Matrix Theory description and show that it leads to a dual description for the behavior at large charge. We use the Random Matrix Theory to prove the existence of the double scaling limit~\eqref{newl}. In section 4 we present some of the detailed predictions of RMT, including an analytic solution for the leading terms in the 't Hooft expansion. From this we extract the exponentially small pieces at strong 't Hooft coupling which describe deviations from effective field theory predictions. 
Finally, in section 5 we study the extremal correlators in Argyres-Douglas rank 1 theories and make some comments on the dual RMT. We compute the first few terms in the large charge expansion and make comparisons with bootstrap data and effective field theory. Some technical details are collected in an appendix.

\section{$\cN=2$ Coulomb Branch Operators}\label{CBoperators}
We begin with a lightening review of the representation theory of $\cN=2$ superconformal theories in four dimensions (a more detailed presentation can be found in \cite{Dolan:2002zh}). The symmetries consist of the usual Poincar\'e and  conformal generators, along with 16 supercharges, usually denoted by $Q_{\,\alpha}^a,\bar Q_{\dot{\alpha}a} $ and $S^a_{\,\alpha},\bar S_{\dot{\alpha}a}$, where $a=1,2$ labels the supercharges and $\alpha, \dot \alpha$ are the usual Lorentz indices. In addition there is $\su(2)_R\times \u(1)_R$ $R$-symmetry. The local operators in the theory furnish representations of this $\su(2,2|2)$ superconformal algebra.

An important class of operators in $\cN=2$ superconformal theories are the so-called Coulomb branch operators. These are particular short representations of $\su(2,2|2)$. 
The defining property of Coulomb branch operators is that these are superconformal primary operators that  are furthermore annihilated by all the $\bar Q_{\dot{\alpha}a} $ supercharges. 

Using unitarity of the representations, it is a standard exercise to show that the dimension of these operators is fixed to be $\Delta=R/2$, where $R$ denotes the $R$-charge under the $\u(1)_R$ R-symmetry. Furthermore, one finds that such operators must be $\su(2)_R$ singlets and have no right-moving spin indices. In fact, in all the constructions relevant to this paper they would have no left-spin indices either and hence the operators would be Lorentz scalars. We will denote these operators by $\cO_I$, where $I$ is an index that labels the different Coulomb branch operators.

\begin{figure}[t]
		\centering 
		\includegraphics[width=0.5\linewidth]{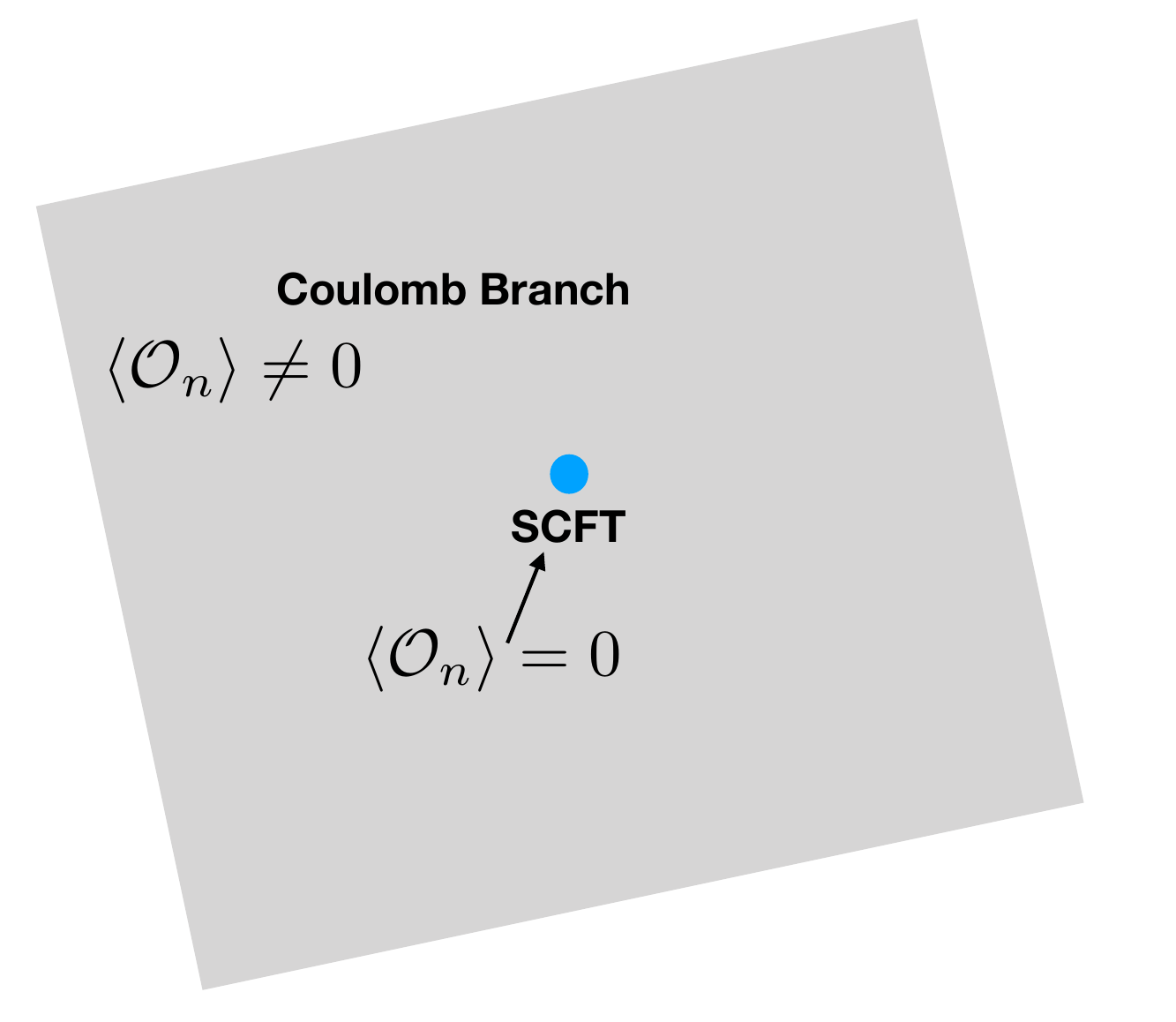}
		\caption{Structure of the moduli space of Coulomb branch vacua. The grey area, where the $\u(1)_R$ and conformal symmetries are spontaneously broken, is parametrized by non-zero VEVs of Coulomb branch operators $\cO_I$.}\label{CBranch}
\end{figure}

Let us now review the significance of these operators. Typically, $\cN=2$ theories have a moduli space of degenerate vacua. A distinguished class of vacua are those where the low-energy theory has a free photon along with its supersymmetric partners.\footnote{Possibly there could be additional degrees of freedom in more complicated models than those we consider here.} These are the Coulomb branch vacua. In these vacua the $\su(2)_R$ symmetry is preserved but the $\u(1)_R$ symmetry and the conformal symmetries are spontaneously broken. More precisely, these vacua preserve the Poincar\'e symmetries along with 8 supercharges and $\su(2)_R$, while the other symmetries in $\su(2,2|2)$ are spontaneously broken and hence non-linearly realized on the vacua. 
These Coulomb branch vacua are parameterized by the Vacuum Expectation Values (VEVs) of the Coulomb Branch operators defined above. Therefore, if we choose to be in a vacuum on the Coulomb branch, the one-point functions of Coulomb branch operators are nonzero $\langle O_I\rangle \neq 0$, see figure \ref{CBranch}. 

An all-important  property of the operators $O_I$ is that if we consider their operator product expansion it only contains regular terms. This leads to a ring structure 
$$\cO_I(x)\cO_J(y)\sim C_{IJ}^K\cO_K(x)+\cdots\ec$$
where the $\cdots$ stand for terms that vanish as $x\to y$.

In the cases studied in this paper the ring will be freely generated by a single operator. These are called rank-1 theories. We will therefore label the operators by $\cO_n$ with $n\geq 0$ an integer (by convention $\cO_0=1$). The operator product expansion then takes the form 
\begin{equation}\label{OPE}\cO_{n}(x)\cO_{m}(y)\sim \cO_{n+m}(x)+\cdots\ed\end{equation}
Of course, one can make various redefinitions of the operators but the choice in~\eqref{OPE} is the most convenient one for our purposes.

If we are at the fixed point (and not on the Coulomb branch) the correlation functions of operators involving just the $\cO_n$ of course vanish by the $\u(1)_R$ symmetry. However, there are non-trivial correlation functions that involve Coulomb branch operators and anti-Coulomb branch operators.

The study of supersymmetric gauge theories in the 90s was mostly restricted to holomorphic observables, namely, a great deal was learned about the correlation functions of the operators $\cO$ away from the fixed point. However, with the progress on supersymmetric localization it became possible to consider more complicated observables. 

For instance, consider the following correlation function measured at the conformal fixed point:
\begin{equation}\label{extremal}\langle \cO_{i_1}(x_1)\cO_{i_2}(x_2)\cdots \cO_{i_n}(x_n)\bar \cO_j(y)\rangle_{SCFT}  \end{equation}
The correlation function~\eqref{extremal}  involves both Coulomb branch operators and exactly one anti-Coulomb branch operator. Using the $\u(1)_R$ selection rule along with $\Delta=R/2$, we see that such a correlation function is potentially non-vanishing only if 
$${\Delta}_{i_1}+\cdots +{\Delta}_{i_n}={\Delta}_j~.$$
Since there is only one anti-Coulomb branch operator in~\eqref{extremal}, such correlation functions are sometimes called ``extremal.'' In some sense, these are the simplest nontrivial correlation functions of the superconformal field theory. In fact, they are probably the simplest nontrivial correlation functions of any four-dimensional gauge theory.

Using the superconformal Ward identities it is possible to show~\cite{Papadodimas:2009eu} that the space dependence of the correlation function~\eqref{extremal} is given by
\begin{equation}\label{extremali}\langle \cO_{i_1}(x_1)\cO_{i_2}(x_2)\cdots \cO_{i_n}(x_n)\bar \cO_j(y)\rangle_{SCFT} = G_{i_1,..,i_n ; j}(\tau,\bar\tau)\prod_{k=1}^n{1\over (y-x_k)^{2i_k\Delta_{\cO} }}~,  \end{equation}
where $G_{i_1,..,i_n ; j}(\tau, \bar \tau)$ is a function of the (complexified) exactly marginal coupling constant $\tau = {\theta \over 2\pi} + {4\pi i\over g_{\textrm{YM}}^2}$ and $\Delta_\cO$ is the dimension of the generator of the ring.
We emphasize that $G_{i_1,..,i_n ; j}$ is non-holomorphic. 

Since there is no singularity in~\eqref{extremali} as $x_i$ approaches some other $x_j$, we can therefore use the OPE~\eqref{OPE} to successively bring the operators $\cO_i$ together and obtain an alternative representation for the function $G_{i_1,..,i_n ; j}(\tau,\bar \tau)$:
\begin{equation}\label{extremalii}\langle \cO_{\sum_{k=1}^n i_k}(x)\bar \cO_j(y)\rangle_{SCFT} = G_{i_1,..,i_n ; j}(\tau,\bar\tau){1\over (y-x)^{2j\Delta_\cO}}~.  \end{equation}
In particular, $G_{i_1,..,i_n ; j}$ is only a function of $\sum_{k=1}^n i_k=j$ and not of the individual $i_k$.
The problem of computing the $G_{i_1,..,i_n ; j}$ is therefore reduced to computing the coupling constant dependence on two-point functions at the SCFT:
\begin{equation}\label{extremaliii}\langle \cO_{j}(x)\bar \cO_j(y)\rangle_{SCFT} = G_{2j}(\tau,\bar\tau){1\over (y-x)^{2j\Delta_\cO}}\ec \end{equation}
where we have defined $G_{2j}\equiv G_{j;j}$ to make contact with previous notation in the literature.

Usually in the conformal bootstrap literature the operators are taken to furnish an orthonormal basis and the nontrivial OPE coefficients are encoded in three-point functions. Here we have instead chosen the OPE coefficients to be trivial~\eqref{OPE} and the nontrivial information is encoded in the two-point functions.

Let us list a few examples (with an increasing level of difficulty) that would be important in this work: 
\begin{itemize}
\item {\bf The free $\cN=2$ $\u(1)$ vector multiplet.} This is a trivial (free) superconformal field theory. The chiral ring is generated by the complex scalar field $\phi$ in the vector multiplet. Hence $\Delta_\cO=1$. The theory has an exactly marginal coupling constant $\tau$  which couples to the operator in the chiral ring $\phi^2$ through $\int d^4\theta\ \tau\phi^2$ (though since the theory is free the conformal field theory data does not depend on $\tau$).
One can compute $G_{2n}$ using Wick contractions and one finds (in a certain convention for the normalization of the generator of the chiral ring -- for now, the most important thing is the $n$-dependence): 
\begin{equation}\label{FV}G_{2n}^{\textrm{Free-Vector}}= n!\,\ed\end{equation}

\item {\bf $\cN=4$ supersymmetric Yang-Mills theory with gauge group $\SU(2)$.} Here the chiral ring is generated by the gauge invariant operator $\Tr(\phi^2)$ where, in the language of $\cN=2$ supersymmetry, $\phi$ is the adjoint scalar in the $\cN=2$ vector multiplet. Hence, $\Delta_\cO=2$ in this model. There is again an exactly marginal coupling $\tau$ which couples to the operator in the chiral ring $\Tr(\phi^2)$ through $\int d^4\theta\ \tau\Tr(\phi^2)$.
The conformal field theory data depends nontrivially on $\tau$.
 Here, surprisingly, in spite of the fact that the model is interacting, there is a non-renormalization theorem \cite{Lee:1998bxa} for the extremal correlation functions~\eqref{extremal}. The statement is that the correlation function is given by its value at tree-level. Thus, (again, in some convenient normalization of the generator of the chiral ring, which allows to focus our attention on the $n$-dependence)
\begin{equation}\label{MSYM}G^{\cN=4}_{2n}=(2n+1)!~.\end{equation} 
 
\item {\bf $\SU(2)$ gauge theory with $N_f =4$ fundamental hypermultiplets.} 
The chiral ring is again generated by the gauge invariant operator $\Tr(\phi^2)$ where $\phi$ is the adjoint scalar in the $\cN=2$ vector multiplet and one has an exactly marginal coupling constant,~$\tau$. Hence, $\Delta_\cO=2$ in this model as well.
The functions $G_{2n}$ have a nontrivial perturbative expansion, instanton corrections, and interesting resurgence properties and $n$ dependence~\cite{Baggio:2014ioa,Baggio:2014sna,Baggio:2015vxa,Gerchkovitz:2016gxx,Aniceto:2014hoa,Honda:2016mvg,Bourget:2018obm,Beccaria:2018xxl}. They are not known in closed form and one of the main aims of this note is to get a better understanding of these correlation functions. 

\item {\bf Rank-1 Argyres-Douglas fixed points.} These superconformal field theories have no exactly marginal coupling constants but that does not mean that the extremal correlation functions are void of content. The structure of the chiral ring~\eqref{OPE} allows to choose the normalization of the generator of the ring at will as long as the higher operators in the tower are appropriately redefined. Hence  $G_{2n}$ can be rescaled by $\lambda^n$ with an arbitrary constant $\lambda$. This means that the ratios $G_{2n}/G_2^n$ are scheme independent and carry intrinsic information about the OPE coefficients of the conformal field theory. These are pure numbers, which would be very interesting to compute. We will address this problem in section \ref{RMTAD} where we also compare our results with some impressive numerical bootstrap data concerning these coefficients \cite{Beem:2014zpa,Lemos:2015awa,Cornagliotto:2017snu}.

\end{itemize}

\subsection{Localization Computations of Extremal Correlators}\label{localized}
A powerful technique to determine (in principle) the functions $G_{2n}$ is through supersymmetric localization. This technique is far more computationally efficient (when it exists) than the ordinary perturbation theory and semi-classical techniques. Any $\cN=2$ SCFT can be placed on $S^4$ while preserving an $\osp(4|2)$ subalgebra\footnote{The superalgebra $\osp(4|2)$ contains a bosonic subalgebra given by $\u(1)_R \times \sp(4)$. The $\u(1)_R$ factor is the Cartan of the original $\su(2)_R$ R-symmetry while the $\sp(4)$ factor gives the isometry of $S^4$. It will be very important below that the original $\u(1)_R$ symmetry is broken on $S^4$.} of the superconformal algebra 
$\su(2|2)$. A systematic procedure to compute the supersymmetric partition function $Z_{S^4}$ of Lagrangian $\cN=2$ theories has been developed in \cite{Pestun:2007rz}. The supersymmetric partition function on $S^4$ can be written as:
\be\label{pfschem}
Z_{S^4}[\tau,\bar \tau] = \int_{\mathfrak{t}} da \,\Delta(a)\, |F(a; \tau,\bar \tau)|^2\ed
\ee
The above integral is performed over a Cartan subalgebra $\mathfrak{t}$ of the gauge group $G$ taking into account an additional factor from the associated Vandermonde determinant $\Delta(a)$. The function $F(a; \tau, \bar \tau)$ has a complicated structure, organized in terms of the instanton partition function, whose explicit details are not important at the moment and will be discussed later in the paper.

Starting from the $S^4$ partition function \eqref{pfschem}, it is  possible to obtain the extremal correlation functions~\eqref{extremaliii}. Below we will explain the procedure for $\SU(2)$ gauge theory with four hypers or $\cN=4$ SYM with gauge group $\SU(2)$. (Slight modifications are necessary in order to study the free vector multiplet, as we will see below.)

\begin{figure}[t]
		\centering 
		\includegraphics[width=0.6\linewidth]{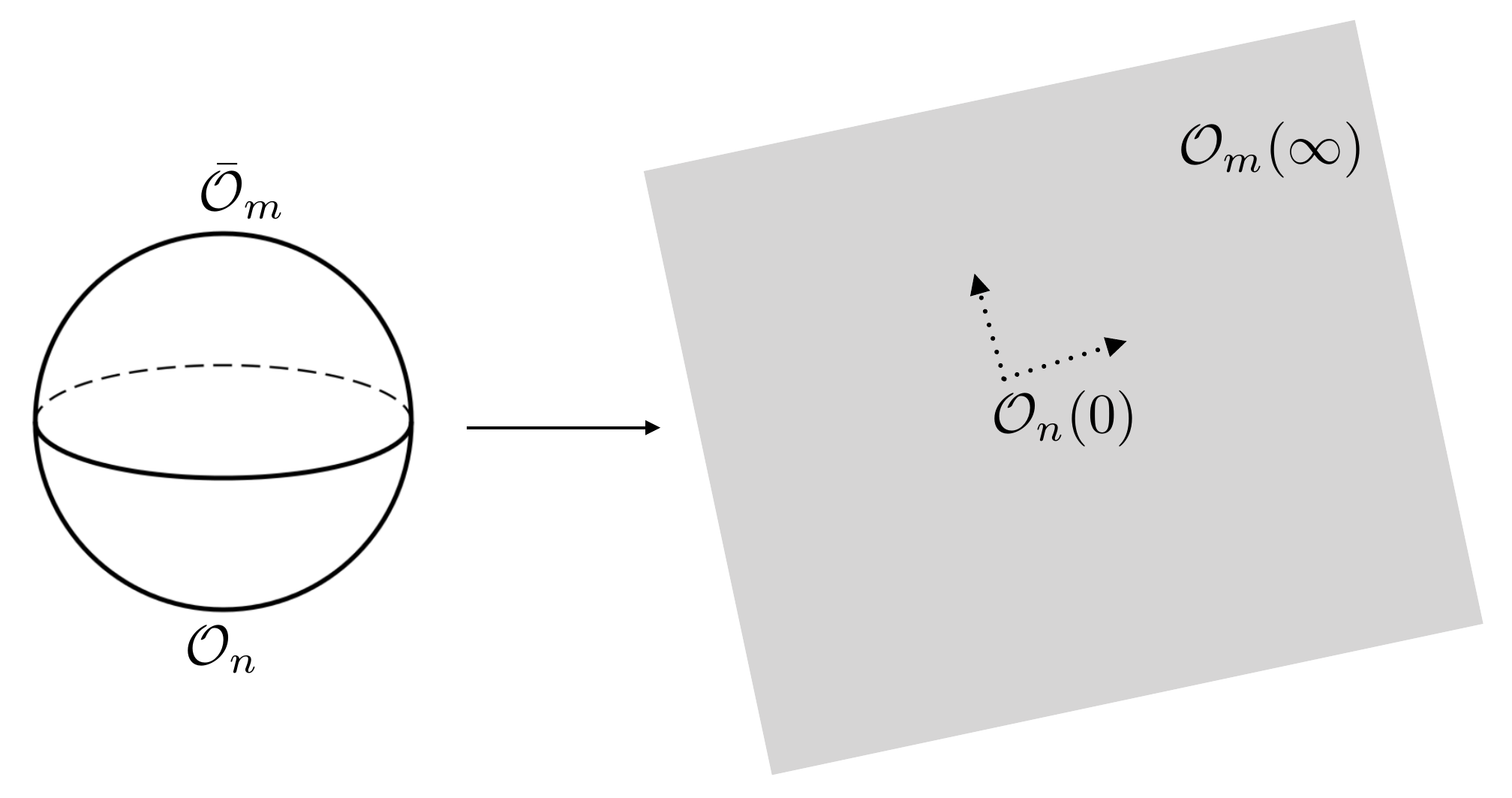}
		\caption{Coulomb branch operators in flat space get mapped to north $(N)$ and south $(S)$ pole insertions on $S^4$.}\label{spheremap}
\end{figure}

First, it is possible to prove \cite{Gerchkovitz:2016gxx} that a derivative of $Z_{S^4}$ with respect to~$\tau$ brings down an insertion of the chiral ring generator $\cO_1(N)$ at the north pole of the sphere and a derivative with respect to $\bar \tau$ bring down an insertion of $\bar{\cO}_1(S)$ at the south pole.\footnote{This holds for $\SU(2)$ gauge theory with four hypers or $\cN=4$ SYM with gauge group $\SU(2)$, but not in the free vector multiplet. A derivation of this fact requires a careful analysis involving the $\cN=2$ supergravity background used in localization. Taking derivatives with respect to the coupling constants $\tau$ and $\bar \tau$ leads to a correlation function of integrated top components. In supersymmetric theories a top component can always be written as $Q^2$ acting on some bottom component; we might think that correlation functions involving integrated top components are completely trivial because we are always free to move around $Q^2$ and get something vanishing. The point is that $Q^2$ on the sphere is not globally well defined, it has a singularity at one of the poles, which if analyzed carefully leads to~\eqref{2pointNS}.} This is depicted in figure \ref{spheremap}.
Therefore,
\be\label{2pointNS}
\langle\cO_1(N)\bar\cO_1(S)\rangle_{S^4} = \frac{\partial^2 \log Z_{S^4}[\tau, \bar\tau]}{\partial\tau\partial\bar\tau}\ed
\ee 

The next crucial point is that even though $\bR^4$ and $S^4$ are conformally equivalent, a direct comparison between the flat space two-point function \eqref{extremaliii} and \eqref{2pointNS} turns out to be far more complicated than expected. One reason for these complications is ultraviolet-sensitive operator mixing. Imagine, for example, the operator $\cO_1$ of dimension $\Delta_{\cO}=2$ together with its associated source $\tau(x)$. Then, consider a new term in the Lagrangian given by $\int dx \sqrt{g}\, \cR(x)\tau(x)$ with $\cR(x)$ the Ricci scalar. This expression is scheme dependent and it leads to a non-vanishing one point function $\langle \cO_1(x)\rangle \neq 0$ in curved space due to mixing of $\cO_1$ and the unit operator $\mathbbm{1}$. Another, not unrelated,  source of complications has to do with generalized trace anomalies. For a discussion of these issues see for instance~\cite{Gomis:2015yaa,Tachikawa:2017aux, Schwimmer:2018hdl,Schwimmer:2019efk,Nakayama:2019mpz}.

In fact, there is no scheme which preserves the  $\osp(4|2)$ subalgebra on  $S^4$  {\it and} removes the operator mixing.

In general, we should always consider mixing between a Coulomb branch operator and a full tower of lower-dimensional operators as follows:
\be\label{mixop}
\cO_n \longrightarrow \cO_n+ \alpha_1(\tau, \bar \tau)\cO_{n-1}R^{-2} + \alpha_2(\tau, \bar \tau)\cO_{n-2}R^{-4} +\dots\ec
\ee
where $R$ is the radius of the round $S^4$. The mixing~\eqref{mixop} violates the $u(1)_R$ symmetry, but it does not violate the supersymmetry that is actually preserved on the sphere, namely $\osp(4|2)$.\footnote{Note that operator mixing with operators of higher dimension is not allowed by locality.}

In order to extract the correlation functions in $\R^4$ that we are ultimately interested in, one therefore has to disentangle the operator mixing which occurs on the sphere~\eqref{mixop}. We now review the prescription \cite{Gerchkovitz:2016gxx} for how to obtain the flat space correlation functions focusing on $\SU(2)$ gauge theory with four hypers or $\cN=4$ SYM with gauge group $\SU(2)$.

Consider the two-point function of chiral ring operators $\cO_n$ which is obtained analogously to \eqref{2pointNS} by:
\be\label{chiralNS}
\langle\cO_n(N)\bar\cO_m(S)\rangle_{S^4} =\frac{1}{Z_{S^4}} \frac{\partial^{n+m} Z_{S^4}[\tau, \bar \tau]}{\partial^n\tau\partial^m\bar\tau}\ec
\ee 
because of operator mixing \eqref{mixop}, this cannot be immediately identified with the corresponding two-point function on flat space. A way to resolve this problem and disentangle the mixing is to treat the two-point function on $S^4$ as an infinite-dimensional matrix:
\be\label{matrixmn}
\cM_{m, n} \equiv \langle\cO_n(N)\bar\cO_m(S)\rangle_{S^4}\ec
\ee 
with $n,m$ non-negative integers and to consider the following ratio of determinants:
\begin{equation*}
\frac{\det{\cM_{(n+1)}}}{\det{\cM_{(n)}}}\ec
\end{equation*}
where $\cM_{(n)}$ denotes an upper-left $n \times n$ sub-matrix of $\cM$. Namely, we divide the determinant of the $(n+1)\times (n+1)$ block by the determinant of the $n\times n$ block. Each block captures the mixing of $\cO_n$ with lower-dimensional operators. As a result, this procedure (which looks quite baroque at first sight) correctly maps the operator insertions from $S^4 $ to $\R^4$.  $G_{2n}$ can be therefore expressed as a ratio of determinants:
\be\label{presc}
G_{2n} = \frac{\det{\cM_{(n+1)}}}{\det{\cM_{(n)}}}\ed
\ee
As an aside, from this general prescription, one can immediately derive an important corollary.
According to~\eqref{presc}, $G_2={1\over Z_{S^4}^2}\left(Z_{S^4}\frac{\partial^2 Z_{S^4}[\tau, \bar \tau]}{\partial\tau\partial\bar\tau}-\frac{\partial Z_{S^4}[\tau, \bar \tau]}{\partial\tau}\frac{\partial Z_{S^4}[\tau, \bar \tau]}{\partial\bar\tau}\right)=\partial\bar\partial \log Z_{S^4}$. On the other hand, since $G_2$ is also, by definition, the Zamolodchikov metric, this implies that~\cite{Gerchkovitz:2014gta,Gomis:2014woa,Gomis:2015yaa,Tachikawa:2017aux,Seiberg:2018ntt}
\be\label{Kahler}
Z_{S^4}[\tau, \bar \tau] = e^{K(\tau, \bar\tau)}\ed
\ee 
Let us now discuss some explicit examples of~\eqref{presc}.
\begin{itemize}
\item {\bf The Free $\cN=2$ $\u(1)$ Vector Multiplet.}
A free $\u(1)$ vector multiplet has an extremely simple $S^4$ partition function given by 
\be\label{FVPF}
Z_{S^4}[\tau,\bar \tau]=\int da e^{-4\pi \Im \tau R^2 a^2}\ec
\ee
where $R$ is the sphere radius.  
The first step in obtaining the $\R^4$ extremal correlation functions consists of computing the matrix \eqref{matrixmn} 
\be
\cM_{k,l}={1\over Z_{S^4}}\int da \,e^{-4\pi \Im \tau R^2 a^2}a^{k+l}\ec
\ee
where $k,l$ range on the non-negative integers. We can always set $R=1$ or modify the formula~\eqref{presc} accordingly such that the radius dependence disappears from the left hand side.
We can now proceed to evaluate the $G_{2n}$. First, we find that for $k+l$ odd $\cM_{k,l}=0 $. For even $k+l$: 
\be
\cM_{k,l}={1\over \sqrt \pi}\left({g_{\rm YM}\over 4\pi}\right)^{k+l} \Gamma\left({k+l\over 2}+{1\over2 }\right)\ed 
\ee
Computing the determinant ratio \eqref{presc} is rather straighforward and leads to, 
\be G_{2n}= \left({g_{\rm YM}\over 4\pi}\right)^{2n} {n!\over 2^{n}}\ed\ee
The only essential part, which is independent of operator normalization, is the $n!$ which perfectly agrees with~\eqref{FV}. All the other factors can be absorbed in normalizing the vacuum partition function and the operators.

\item{\bf $\cN=4$ Supersymmetric Yang-Mills Theory with Gauge Group $\SU(2)$.}
The $S^4$ partition function is rather similar to~\eqref{FVPF}. The only difference here is the Vandermonde factor from the measure over the Coulomb branch as in \eqref{pfschem}:
\be Z_{S^4}[\tau,\bar \tau]=\int da\, (2a^2) e^{-4\pi \Im \tau R^2 a^2}\ed\ee
As previously, in order to extract the physical correlators on $\R^4$ one is instructed to first compute 
\be
\cM_{k,l}={1\over Z_{S^4}}\int da (2a^2) e^{-4\pi \Im \tau R^2 a^2}a^{2k+2l}={2\over \sqrt\pi}\left({g_{\rm YM}^2\over 16\pi^2}\right)^{k+l}\Gamma\left(k+l+{3\over 2}\right)\ec
\ee
and then extract the flat space correlation function from the ratio of determinants~\eqref{presc}.
The ratio of determinants, up to inessential factors that can be absorbed in the normalization of the unit operator and the chiral ring generator, is given by:
\be\label{maxsym}
G_{2n} ={1\over 4^{n}}\left({g_{\rm YM}^2\over 16 \pi^2}\right)^{2n}\!{(2n+1)!}\,\ed
\ee
Hence, the correlation function in $\R^4$ scales like $(2n+1)!\,$, in exact agreement with the explicit tree-level computation in flat space~\eqref{MSYM}.
\end{itemize}

\subsubsection {$\SU(2)$ Gauge Theory with $N_f =4$}\label{gaugeNf4}
We will now discuss our first truly non-trivial example, namely $\SU(2)$ gauge theory coupled to four massless hypermultiplets in the fundamental representation.
The supersymmetric partition function on a round $S^4$ of  radius $R$, which was already introduced in \eqref{pfschem}, can be now written more explicitly as:
\be\label{pestun}
Z_{S^4}[\tau, \bar \tau] = \int_{\bR} da \,e^{-4\pi \Im\tau { a^2} R^2} (2a^2) Z_{\textrm{1-Loop}}(a,R)|Z_\textrm{Inst}(ia, q, R)|^2\ed
\ee
In this expression $Z_{\textrm{1-Loop}}(a,R)$ denotes the one-loop determinant which can be written as a product of Barnes $G$ gamma functions
\be\label{1loop}
Z_{\textrm{1-Loop}}(a,R) = \frac{H(2iaR)H(-2iaR)}{|H(-iaR)H(iaR)|^4}\ec
\ee
with  $H(x) \equiv \re^{(1+\gamma) x^2} G(1+x)G(1-x)$, $\gamma$ being the Euler constant.  The remaining factor in \eqref{pestun} is given by the instanton partition function with $\Omega$-background parameters $\epsilon_1=\epsilon_2 =  {1\over R}$ 
and gauge instanton fugacity $q \equiv e^{2\pi  i\tau}$. 
A comprehensive treatment of the instanton partition function can be found in \cite{Moore:1997dj, Nekrasov:2002qd, Nekrasov:2003rj}. The expansion in the gauge instanton fugacity $q$ grows quickly in complexity, for example up to 2-instantons\footnote{In the context of AGT \eqref{inst} is given in terms of the $U(2)$ instanton partition function which differs from the $SU(2)$ case by the well known $U(1)$ factor \cite{Alday:2009aq}. At the level of extremal correlators this factor is not important since it is equivalent to a K{\"a}hler transformation as discussed in \cite{Gerchkovitz:2016gxx}. We thank S. Hellerman for a useful discussion on the subject.}:
\be \label{inst}
Z_{\inst}(ia, q, R) = 1 + \half e^{2\pi i \tau}(a^2R^2- 3) + {1\over 4}e^{4\pi i \tau}\,\frac{8a^8R^8+a^6R^6 -91a^4R^4 -60a^2R^2+132}{(4a^2R^2+9)^2} +\dots
\ee
Due to the nontrivial ``one-loop'' factor in~\eqref{pestun} and the nontrivial expansion in the instanton fugacity $q$, the resulting extremal correlators in $\R^4$ have a nontrivial perturbative expansion (which matches the standard Feynman perturbation theory) as well as non-perturbative terms. Once again, in order to obtain the $\R^4$ two point functions $G_{2n}$ from~\eqref{pestun} one is instructed to consider the matrix of derivatives \be\label{matrixdef}\cM_{k,l}= {1\over Z_{S^4}}{\partial^k\over \partial \tau^k}{ \partial^l\over \partial\bar\tau^l}Z_{S^4}[\tau,\bar \tau]~,\quad k,l=0,1...
\ed\ee
Notice that if we ignore the instantons altogether the $\tau,\bar\tau$ derivatives in~\eqref{matrixdef} simply lead to insertions of $a^2$. 

It is impractical to hope that we can obtain exact explicit expression for the $G_{2n}$, but it is rather easy to write down the perturbative expansion and also the  perturbative expansion around the first few instanton sectors. For instance, the peturbative expansion of $G_2$ in the zero-instanton sector is
\be
G_2={3\over 8 (\Im\tau)^2}-{135\zeta(3)\over 32\pi^2}{1\over (\Im \tau)^4}+{1575\zeta(5)\over 64\pi^3}{1\over (\Im \tau)^5}+\cdots~.
\ee
 The coefficients quickly grow as one expects from ordinary perturbation theory.
While one can glean a lot of information from the explicit expansion in perturbation theory, as we explained in the introduction, the main focus of this paper is the dependence on $n$ of $G_{2n}$ and the interplay with perturbation theory. This is hard to understand systematically by expanding in $g_{\rm YM}$ and hence we will soon introduce a dual description.

\subsubsection{Large $a$ Expansion}\label{largeradius}

So far we discussed the expansion of~\eqref{pestun} in the gauge coupling. 
Another useful expansion that will be important below is the expansion at large $a$ of the integrand of~\eqref{pestun}. It is clear that this expansion is relevant for the large $n$ limit since in the introduction we argued that the Coulomb branch field is kicked out of the origin by $\phi\sim \sqrt n$. The VEV of the Coulomb branch field is essentially the same as $a$ and hence we will have to study the large $a$ expansion of~\eqref{pestun}. 
While this intuition is essentially correct we will see that there are various subtleties. First, some important effects at large $n$ do not come from large $a$. Second, in the double scaling limit that we explained in the introduction one in fact cannot perform the large $a$ expansion (this is essentially because the field $a$ is kicked out to $\sim \sqrt {n/\Im \tau}\sim \sqrt \lambda$ which may or may not be large).

Since the theory is conformal the integrand in~\eqref{pestun} essentially only depends on $aR$. Thanks to the relationship $\epsilon_1=\epsilon_2={1\over R}$, which holds on the round $S^4$, the large radius expansion of~\eqref{pestun} is tantamount to the standard expansion at small $\epsilon_{1,2}$. This is analogous to the genus expansion of topological string theory, see for instance  \cite{Klemm:2002pa}. We can write in general the four-sphere partition function as:
\be\label{largeR}
Z_{S^4}[\tau, \bar \tau] = \int_{\bR} da \,e^{-4\pi \Im\tau { a^2} R^2} (2a^2)\exp{\biggl(\sum_{g\geq 0}R^{2-2g}a^{2-2g}\left(\cF_g(q)+{ \overline {\cF_g(q)}}\right) \biggr)} \ec
\ee  
where at each order the quantity $\cF_g$ contains contributions both from $Z_{\textrm{1-Loop}}$ and $Z_{\textrm{Inst}}$. (The case $g=1$ should be dealt with separately  since it may have logarithms.)

For the $\SU(2)$ theory with $N_f=4$ massless fundamentals, using \eqref{1loop} and \eqref{inst} we record here the genus 0 and genus 1 cases:
\be\label{f0f1i}
\begin{split} 
\cF_0&=-2  \log (4)+\frac{ q }{2}+\frac{13 q ^2}{64}+\frac{23 q ^3}{192}+\frac{2701 q ^4}{32768}+O\left(q ^5\right)\ec\\
\cF_1&=\frac{1}{6} (3 \log (a )+36 \log(\gamma_G)-3-\log (2))-\frac{3 q }{2}-\frac{23 q ^2}{32}-\frac{15 q ^3}{32}-\frac{2839 q ^4}{8192}+O\left(q ^5\right)\ec\\
\end{split}
\ee
where $\gamma_G$ is the Glaisher constant. 


\subsubsection{Toda Equations}\label{Todaconstraint}
\begin{figure}[h!]
		\centering 
		\includegraphics[width=0.7\linewidth]{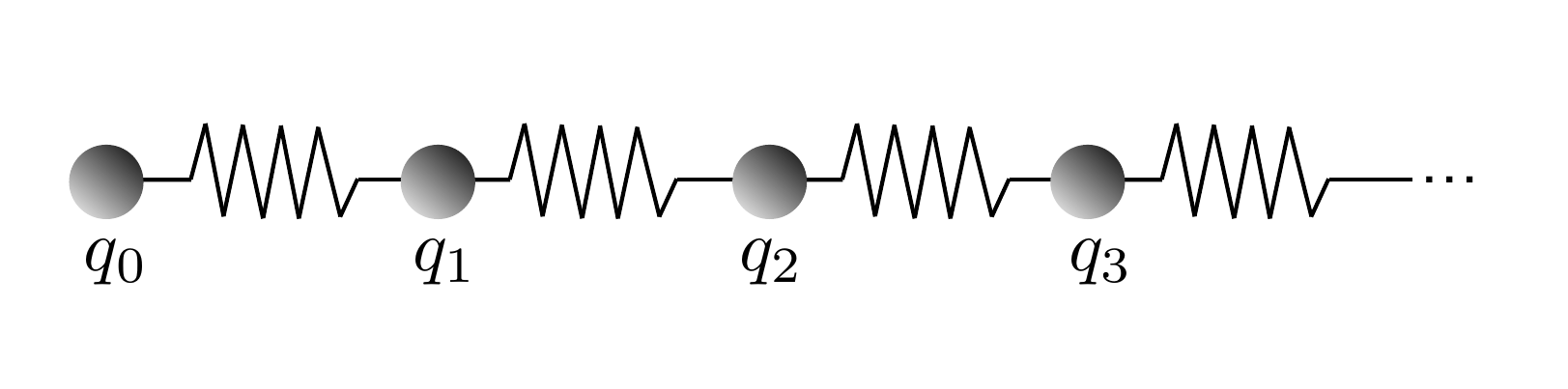}
		\caption{A simple graphic depiction of a half-infinite Toda chain for the evolution of extremal correlators $q_i$'s. Each site can be thought of as an oscillator connected to a neighboring site by a spring. The interaction between each site is governed by the right-hand side of equation \eqref{TODAii}, while the second equation \eqref{TODAiii} sets the boundary condition for the problem.}\label{TodaChainNew}
\end{figure}

The extremal correlation functions $G_{2n}$ satisfy an important, and perhaps surprising, non-perturbative property related to classical integrability. In fact it has been appreciated for a long time that ratios of determinants such as the one appearing in \eqref{presc} are solutions of integrable models. For the rank 1 $\cN=2$ SCFTs theories of interest here, it turns out that the relevant system to consider is a semi-infinite (or open) Toda chain \cite{Hirota1988}. The differential equation obeyed by the two-point functions is given by 
\be\label{TODAi}\partial_\tau\partial_{\bar \tau} \log G_{2n} = {G_{2n+2}\over G_{2n}}-{G_{2n}\over G_{2n-2}}-G_2~,\quad n=1,2...,\ee
where we take $G_0=1$. This can be recast in the more familiar form of a Toda chain by performing a change of variables $G_{2n}\equiv e^{q_n-\log Z[S^4]}$. With this definition $q_0=\log Z[S^4]$ and we obtain the equations 
\begin{align}
\label{TODAii}\partial_\tau\partial_{\bar \tau} q_{n} &= e^{q_{n+1}-q_n}-e^{q_n-q_{n-1}}~,\quad n=1,2...,\\
\label{TODAiii}\partial_\tau\partial_{\bar \tau} q_{0} &= e^{q_1-q_0}\ed
\end{align}
These equations describe a semi-infinite Toda chain of coupled oscillators $q_i$'s shown in figure \ref{TodaChainNew}, where the first oscillator $q_0$ has a prescribed $\tau$ dependence given by the $S^4$ partition function. The same result can be also derived directly in $\R^4$ \cite{Papadodimas:2009eu, Baggio:2014ioa, Baggio:2015vxa} using a four-dimensional version of the $tt^*$-equations \cite{Cecotti:1991me}.  As we will see, the Toda equations provide useful constraints on the large $n$ limit. 

\subsection{Effective Field Theory Predictions}\label{eftpred}
An alternative approach to the large charge limit utilizes the tools of effective field theory (EFT) around such states. Since at large charge the Coulomb branch field is kicked far away from the origin, the physics can be well described by an effective theory for the $\cN=2$ vector superfield $\Psi$. As explained in the introduction, one has to carefully distinguish the limit of large charge at fixed coupling from the limit of large charge with fixed $g_{\rm YM}^2n$.
Since the masses of the (electric) BPS resonances scale like $\sqrt{g_{\rm YM}^2n}$ the effective field theory only applies in the limit of fixed coupling and large charge (or in the limit of large $ g_{\rm YM}^2n$ but not at generic $g_{\rm YM}^2n$). 
In other words, the predictions below apply when the charge $n$ is much larger than any other parameter. 

The authors \cite{Hellerman:2017sur, Hellerman:2018xpi} argued that the effective field theory for $\Psi$ in flat space has two $F$-terms and the rest are $D$-terms: 
\be\label{eftschem}
\scrL_{EFT} = \scrL_{\textrm{free}} + \scrL_{WZ} + \textrm{(D-terms)}~,
\ee
where $\scrL_{\textrm{free}}$ is just the standard kinetic Lagrangian while the leading interaction term $\scrL_{\textrm{WZ}}$ was written down for instance in \cite{Henningson:1995eh, deWit:1996kc, Dine:1997nq} and it is given by:
\be\label{WZ}
\scrL_{\textrm{WZ}} = \alpha \int d^8\theta \log(\Psi)\log(\Psi^\dagger)\ed
\ee
This term is a little tricky because it is written here as a $D$-term but in fact it is an $F$-term. The subscript WZ has been chosen because the bosonic part of this term contains the Wess-Zumino term for the Weyl $a$-anomaly which was used in \cite{Komargodski:2011vj, Komargodski:2011xv} to prove the a-theorem in four dimensions. From the trace anomaly matching in the proof of the $a$-theorem we learn that $\alpha$ captures the difference between the $a$-anomaly of the SCFT in the ultraviolet and the free vector on the Coulomb branch. For convenience below is a list of all the values of $\alpha$ which are relevant for the theories studied in this paper:
\smallskip
\renewcommand{\arraystretch}{1.6}
\renewcommand\tabcolsep{6pt}
\begin{table}[H]
  \centering
  \begin{tabular}{ |c|c|c|c| }
\hline
{\bf Theory} &  {$a_{\textrm{CFT}}$} &  {$a_{\textrm{free}}$} & $\alpha \equiv 2(a_{\textrm{CFT}} - a_{\textrm{free}})$  \\
\hline
\hline
$\u(1)$ Free Vector &  ${5\over 24}$ & ${5\over 24}$ & $0$  \\
\hline
$\cN=4$, $\SU(2)$ SYM &  ${3 \over 4}$ & ${1\over 4}$ & $1$  \\
\hline
$\cN=2, \SU(2)\, N_f=4$&  ${23 \over 24}$&  ${5\over 24}$ & ${3\over 2}$ \\
\hline
$AD_{N_f=1}(\SU(2))$&  ${43 \over 120}$&  ${5\over 24}$ & ${3\over 10}$ \\
\hline
$AD_{N_f=2}(\SU(2))$&  ${11 \over 24}$&  ${5\over 24}$ & ${1\over 2}$ \\
\hline
$AD_{N_f=3}(\SU(2))$&  ${14 \over 24}$&  ${5\over 24}$ & ${3\over 4}$ \\
\hline
\end{tabular}
  \caption{A list of values for the coefficient $\alpha$ of the Wess-Zumino term in the effective theory for the conformal symmetry breaking dilaton. On the last three rows we use the notation $AD_{N_f=1,2,3}(\SU(2))$ to denote the Argyres-Douglas IR superconformal fixed point realized from a $\cN=2$ $\SU(2)$ theory with $N_f=1,2,3$ fundamental hypermultiplets.}
  \label{tab:alphavalues}
\end{table}

The computation of extremal correlation functions is insensitive to the $D$-terms in~\eqref{eftschem}. As a result, for the extremal correlators the effective theory is extremely useful as long as the charge is the largest parameter. One can hope to compute all the power corrections in $1/n$ since there are no counter-$F$-terms. 
It is convenient to express the predictions for $G_{2n}$ in terms of the $q_n$ variables, which, again, are defined by $G_{2n}\equiv e^{q_n-\log Z[S^4]}$. 
This analysis has been carried out in \cite{Hellerman:2018xpi} where the authors found that (using the Toda equations along with some effective field theory computations) 
\be\label{EFTpred}q_n = A(\tau,\bar \tau)n+B(\tau,\bar \tau)+\log\Gamma(n\Delta_{\cO}+\alpha+1)+\cdots~,\ee 
where the $\cdots$ stands for terms which are exponentially small in the large $n$ limit. The functions $A,B$ are not determined by effective field theory -- in fact, they depend on how the generator of the chiral ring is normalized and how the VEV of the unit operator is normalized. 

A striking fact is that except for the term linear in $n$ and the term of order $n^0$, all the analytic terms in the expansion around $n=\infty$ are independent of the coupling constant $\tau$. We will see that this is not the case in the double scaling limit.

\section{Extremal Correlators from Random Matrix Theory}\label{extrandom}

\subsection{Emergence of Random Matrix Theory}\label{sqcdrm}
It is useful to examine in detail the zero-instanton sector of the extremal correlators. Namely, we begin with the $S^4$ partition function omitting the instanton contributions 
\be\label{pestunper}
Z^{\textrm{pert}}_{S^4}[\tau, \bar \tau] = \int_{\bR} da \,e^{-4\pi \Im\tau a^2} (2a^{2}) Z_{\textrm{1-Loop}}(a)\ec
\ee
where, as before, 
\be\label{1loopp}
Z_{\textrm{1-Loop}}(a) = \frac{H(2ia)H(-2ia)}{|H(-ia)H(ia)|^4}~.
\ee
The superscript ``pert" in~\eqref{pestunper} stands for the fact that this expression for the $S^4$ partition function is valid to all orders in perturbation theory but does not capture the gauge theory instanton contributions (we will discuss later whether this omission is consequential).

The algorithm~\eqref{matrixdef} now simplifies since the $\tau$ derivatives lead to insertions of powers of $a^2$, so that we only need to consider the matrix (up to some unimportant constants which can be absorbed in the normalization of the chiral ring generator)
\be\label{matrix2}
(\cM)_{k,l}= {2\over Z^{\textrm{pert}}_{S^4}}\int_{\bR} da \,e^{-4\pi \Im\tau a^2} a^{2+2k+2l} Z_{\textrm{1-Loop}}(a)~.
\ee
We are then instructed to compute the ratio of determinants~\eqref{presc} from which we finally obtain the all-orders in perturbation theory result for the extremal correlators. In what follows we drop the prefactor $2/Z^{\textrm{pert}}_{S^4}$ in~\eqref{matrix2} since it is easy re-instate it eventually when necessary. Therefore, for what follows
\be\label{matrix}(\cM)_{k,l}=\int_{\bR} da \,e^{-4\pi \Im\tau a^2} a^{2+2k+2l} Z_{\textrm{1-Loop}}(a)~.\ee
Note that while the expansion in the coupling constant is only asymptotic, the integrals~\eqref{matrix} are convergent and hence lead to a well-defined answer.

We are interested in the large $n$ limit of the extremal correlators $G_{2n}$, which involves taking ratios of increasingly large determinants. A convenient way to think about this problem is through the Andr\'eief identify (sometimes also called the Gram or Heine identity), which can be stated in generality as follows: Given two sets of $N$ functions $\{f_k(y);\,g_k(y)\}^{N-1}_{k=0}$ and a measure $d\mu(y)$ we have 
\be
\int \prod_{j=0}^{N-1} d\mu(y_j) \det_{ab}(f_a(y_b))\det_{cd}(g_c(y_d))=N!\det_{ab} \int d\mu(y)f_a(y)g_b(y)
\ee
where $a,b,c,d$ all range on $0,\dots, N-1$. Roughly speaking, the identity relates a determinant of integrals to a multivariate integral over determinants. 

\noindent This identity can be readily applied to~\eqref{matrix} by identifying $d\mu(y)\leftrightarrow da e^{-4\pi \Im\tau a^2}a^2Z_{\textrm{1-Loop}}(a)$, and by identifying $f_k(y)\leftrightarrow a^{2k}$ and $g_k(y)\leftrightarrow a^{2k}$. Another useful identity is the standard Vandermonde determinant 
\be\det_{ab}(y_a^b)\det_{cd}(y_c^d) = \prod_{j<k}(y_j-y_k)^2~.\ee
Using all these ingredients we can then rewrite the determinants of~\eqref{matrix} as 
\be\label{comesRMT}\det \cM_{(n)}={1\over n!}\int \prod_{j=0}^{n-1}dy_j e^{-4\pi \Im\tau y_j^2}y_j^2Z_{\textrm{1-Loop}}(y_j)  \prod_{j<k}(y^2_j-y^2_k)^2~. \ee
The integrals over the variables $y_j$ are over $(-\infty,\infty)$. It is convenient to convert them into integrals over the half line $y_j^2=x_j$, in terms of which we obtain  
\be\label{comesRMTi}\det\cM_{(n)}={1\over n!}\int_0^\infty \prod_{j=0}^{n-1}dx_j e^{-4\pi \Im\tau x_j}x_j^{\half}Z_{\textrm{1-Loop}}\left(\sqrt{x_j}\right)  \prod_{j<k}(x_j-x_k)^2~. \ee
This form has the advantage that the factor $\prod_{j=0}^{n-1}(x_j-x_k)^2$ is readily recognized as the usual repulsion terms between the eigenvalues of a matrix.

The problem of computing the $n\times n$ determinants, and hence the extremal correlators, therefore reduces naturally to the problem of random $n\times n$ matrices with the ensemble weight determined by the {\it integrand} of the four-sphere partition function. 
In the original problem of computing the four-sphere partition function and the extremal correlators we had a one-dimensional integral over the physical Coulomb branch. In equation~\eqref{comesRMTi} we have all of a sudden $n$ random variables. We can think of them physically as the elementary quanta that make the states created by $(\Tr(\phi^2))^n$. 

Since the integral in~\eqref{comesRMTi} is over a half line, the ensemble of random matrices that we are dealing with is not the standard ensemble of Hermitian matrices but of the so-called Wishart matrices,\footnote{The ensemble is also sometimes referred to as the Altland-Zirnbauer ensemble \cite{Altland:1997zz}.} which are given by $W=HH^\dagger$ with $H$ a general complex $n\times M$ matrix with $M\geq n$. 

Let us quickly review some important properties of the ensemble. For a comprehensive review and references see~\cite{2017arXiv171207903L}. We begin with an integral over the general complex  $n\times M$ matrix $H$ with $M\geq n$. We postulate bi-fundamental {\it emergent} gauge symmetry (which is in accord with the type of models we encounter here) under $H\to UHV$ with unitary $U,V$. The integral therefore takes the form  
\be
\int[dH] F(\Tr HH^\dagger,\Tr (HH^\dagger)^2,...) \ee 
with $F$ an arbitrary potential function which is only a function of the eigenvalues of $HH^\dagger$ and hence compatible with the emergent bi-fundamental gauge symmetry. First we would like to change variables to an integral over $W=HH^\dagger$, where $W$ is a Hermitian non-negative $n\times n$ matrix. 
This change of variables is reviewed in \cite{2017arXiv171207903L} and results in an integral over Hermitian semi-positive definite matrices $W$ 
\be\label{PIW} \int[dW] (\det W)^{M-n} F(\Tr W,\Tr W^2,...)~.
\ee 
We can then change variables to the eigenvalues of $W$ and obtain 
$n$ integrals from $0$ to $\infty$ with the Vandermonde interaction term between the eigenvalues:
\be \int \prod_{i=1}^ndw_i \prod_i w_i^{M-n} F(\sum w_i,\sum w_i^2,...) \prod_{j<k}(w_j-w_k)^2~.
\ee

The preceding discussion allows to realize the expression~\eqref{comesRMTi} in terms of an ensemble of Wishart matrices. An important question is how to choose $M$. In~\eqref{PIW} we see that we can absorb the determinant into the potential over the eigenvalues by simply writing $\det W = e^{\Tr \log W}$. For this reason we will choose $M=n$.\footnote{In some situations it might be more natural that the determinant would arise from the measure rather than from the potential for $H$ -- in this paper we just choose $M=n$.}

Comparing with~\eqref{comesRMTi} we can now identify the potential for the semi-positive definite Hermitian matrices $W$ and rewrite the matrix model as an integral over $W$ with the potential 
\be\label{potential}e^{-V}=e^{-4\pi \Im \tau \Tr W +\frac12 \Tr \log W+ \Tr \log Z_{\textrm{1-Loop}} (\sqrt  W) }~.\ee 
It will be important later that this potential is has a single-trace structure. This guarantees that the only interaction among the eigenvalues is due to the Vandermonde factor.

In order to warm up and understand the physical consequences of this matrix model we will now solve a simplified version of it and explain why this simplified version adequately describes the large $n$ limit. 

\subsection{A Large $n$ Approximation of the Random Matrix Model}\label{largenpreds}

Starting from the potential for the eigenvalues~\eqref{potential} we will now make a certain simplification and keep only the first two leading terms for large eigenvalues of $W$. We will then justify this assumption by self-consistency and argue that this adequately describes the leading and sub-leading effects in the large $n$ limit. Indeed, we previously argued that the large $n$ limit and the large eigenvalue limit should be related. Here we will see how this is borne out.

It is easy to check from the explicit expression for  $\log Z_{\textrm{1-Loop}}$ that at large $W$ it becomes $\half \Tr \log W$ and hence our simplified matrix model has the potential 
\be\label{potentialsim}e^{-V}=e^{-4\pi \Im \tau \Tr W +\Tr \log W}~.\ee 
In addition, in $\log Z_{\textrm{1-Loop}}$ there is a piece quadratic in $a^2$, which can be re-absorbed by shifting $\Im \tau$ by a constant.
This approximation that we are doing amounts to retaining only $\cF_0$ and $\cF_1$ in the general expansion of~\eqref{largeR}.

In terms of the eigenvalues, the ensemble~\eqref{potentialsim} reduces to studying the integrals: 
\be
\label{comesRMTii}\det\cM_{(n)}={{{\re}^{12 n \log \gamma_G-n-{n\over 3}\log 2}}\over n!}\int_0^\infty \prod_{j=0}^{n-1}dx_j e^{-4\pi \Im\tau_{\rm R} x_j}x_j  \prod_{j<k}(x_j-x_k)^2~,
\ee
where \be \Im\tau_{ \rm R} \equiv \Im\tau +{1\over 2\pi}\log 2\ed \ee
These integrals are solvable -- they are essentially the normalization factor of the $\upbeta$-Wishart-Laguerre ensemble. In general \cite{szego1959orthogonal} (we follow the notation of \cite{Kumar2017}), if \be\label{cwl}1=C_{n,\upalpha,\upbeta} \int_0^\infty \prod_{i=1}^n d\lambda_i  \prod_{j<k}|\lambda_j-\lambda_k|^\upbeta\prod_{j=1}^n \lambda_j^\upalpha e^{-\upbeta/2\sum_{j=1}^n\lambda_j}\ec\ee
then 
\be\label{normRMT}
C_{n,\upalpha,\upbeta}=\left({\upbeta\over 2}\right)^\upgamma\ \prod_{j=0}^{n-1} {\Gamma\left(\upbeta/2+1\right)\over \Gamma\left({\upbeta\over 2}(j+1) +1\right)\Gamma\left({\upbeta\over 2} j+\upalpha+1\right)}~, \ee
where $\gamma=n\left(\upalpha+{\upbeta\over 2} (n-1)+1\right)$.
Our interest is in the case of $\upalpha=1$, $\upbeta=2$, for which we find that 
\be C_{n,1,2}= \left(\prod_{j=0}^{n-1} {1\over \Gamma\left(j+2\right)}\right)^2 \ed\ee
This result readily applies to our problem after a rescaling of the variables. We thus find that
\be\label{approxZ}
\begin{split}
\det \cM_{(n)}&={{{\re}^{12 n \log \gamma_G-n-{n\over 3}\log 2}}\over n!}\int_0^\infty \prod_{j=0}^{n-1}dy_j e^{-4\pi \Im\tau_R y_j}y_j  \prod_{j<k}(y_j-y_k)^2\\
&={{{\re}^{12 n \log \gamma_G-n-{n\over 3}\log 2}} \over (4\pi \Im\tau_R)^{n(n+1)}}{1\over n!}\left( \prod_{j=0}^{n-1}  \Gamma\left(j+2\right) \right)^2~.
\end{split}
\ee
From this we can obtain an approximation (which is yet to be justified) for the extremal correlators \eqref{presc}  
\begin{equation}\begin{split}\label{almostthere} \log G_{2n}&=(-2n-2)\log (4\pi \Im\tau_R)+ \log {\Gamma^2(n+2)\over n+1}+ { {12 \log \gamma_G-1-{1\over 3}\log 2}}\\&= 2n\log n +2\log n-2(n+1)\log (4\pi \Im\tau_R)-2n+\log 2\pi \\
 &+ {{12 \log \gamma_G-1-{1\over 3}\log 2}}+{7\over 6n}+\cdots~,\end{split}\end{equation}
where we used Stirling's formula.\footnote{We remind that the first few terms are $$\log \Gamma(z)\sim z\log z -z-{1\over 2}\log\left({z\over 2\pi}\right)+{1\over 12z}+\cdots~.$$ } 

Let us now justify why we should trust the first few terms in~\eqref{almostthere}. The idea is to consider the eigenvalue distribution for the matrix model with potential~\eqref{potentialsim}. The eigenvalue distribution function is known as the Mar\v{c}enko-Pastur distribution \cite{Marcenko1967} and it takes the form
 \be\label{MP}\rho(x)={1\over 2\pi x }\sqrt{\left({4\pi \Im\tau_R \,x\over n}-a\right)\left(b-{4\pi \Im\tau_R \,x\over n}\right)}~,\ee
with lower edge $a$ and upper edge $b$ respectively given by 
\be
a=\left(-1+\sqrt{\frac{1}{n}+1}\,\right)^2={1\over 4n^2} +\cdots\ec \qquad b=\left(1+\sqrt{\frac{1}{n}+1}\,\right)^2=4+\cdots\ed 
\ee
Consequently, at large $n$ the eigenvalues range from essentially zero to eigenvalues of order $n/\Im\tau$. As a result, the typical eigenvalue is of order $n/\Im\tau$. 
If we keep $\Im\tau$ fixed and take $n\to\infty$ we therefore see that the typical eigenvalue is large, of order of the size of the matrix. This retroactively justifies the approximation~\eqref{potentialsim}, where we omitted terms that are small when the eigenvalues are large. Two general comments: 

\begin{itemize} 
\item It is clear from the above discussion that in the double scaling limit 
\be  n, \Im \tau  \to \infty, \qquad \lambda={n\over \Im \tau} \quad {\rm fixed}\ec\ee
the eigenvalues are of the order of the coupling $\lambda$ and they are thus fixed as $n\to \infty$.  Hence, one cannot utilize the above expansion in large eigenvalues. As a result, the large $n$ expansion in the double scaling limit is much richer than~\eqref{almostthere}. 

\item One should not expect that \eqref{almostthere} is correct to all orders in $1/n$ since the minority of eigenvalues that are not of order $n$ could modify the answer~\eqref{almostthere}. Indeed, the effective field theory prediction as reviewed in section \ref{eftpred} (we substitute the relevant value of $\alpha$ from table \ref{tab:alphavalues} into \eqref{EFTpred}) is
\be\label{expandedEFT}\log G^{\textrm{EFT}}_{2n}=\log  \Gamma(2n+5/2)=2n\log 2n+2\log 2n-2n+\frac12\log 2\pi +{47\over 48n}+\cdots\ed\ee
As explained around~\eqref{EFTpred}, the effective field theory does not predict terms linear and constant in $n$, hence, one should not trust these terms in~\eqref{expandedEFT}.  
Comparing~\eqref{almostthere} and \eqref{expandedEFT} we see that the coefficients of $n\log n$ 
and $\log n$ exactly match! 
However, the $1/n$ term disagrees and we will momentarily see that this is due to the improbable small eigenvalues.  
\end{itemize}

\subsection{The RMT$\leftrightarrow$ NG-Boson Duality at Higher Order}

In the analysis around~\eqref{almostthere} the coefficients at order $\cO(n \log n)$ and $\cO(\log n)$ have been correctly determined (and in agreement with EFT) using the genus expansion (including contributions from $\cF_0$ and $\cF_1$ only) of the supersymmetric partition function which then led to the ensemble~\eqref{potentialsim}. 

In this subsection we will argue that to capture correctly the contribution of order $\cO\left({1\over n}\right)$ we must take into account the behavior of the potential \eqref{potential} for small eigenvalues. The behavior of the integrand for very small $x$ in~\eqref{comesRMTi} could be thought of as ``$\cF_{\infty}$" in terms of the genus expansion. In the approximation~\eqref{comesRMTii} the leading contribution to the potential for the eigenvalues near the origin is $\log x$. However, in the full random matrix theory~\eqref{comesRMTi} the potential near the origin is $\frac12\log x$. We can think about it as a singular perturbation of the ensemble~\eqref{potentialsim}.

In the RMT literature there has been extensive study on perturbation of the Wishart-Laguerre ensemble, see for instance \cite{hank, hank2, Chen_1998, Chen1}. Let us now describe how to determine the expansion coefficient of $G_{2n}$ at order $\cO\left({1\over n}\right)$ using these techniques. First, it is useful to rewrite \eqref{comesRMTi} as (we rescaled the integration variable)
\be\label{detM}
\det \cM_{(n)} =  { {{\re}^{ 12 n \log \gamma_G-n-{n \over 3}\log 2}}\over{\left({ 4 \pi \Im \tau_R}\right)^{n(n+1)}}n!} \int_{\bR^+} d^n x \prod_{i<j}(x_i-x_j)^2 \prod_{i=1}^n e^{-x_i}x_i^{\half}(x_i+{4 \pi}\,{\Im \tau_R})^{\half}\,U(x_i, { \Im \tau_R})\ed 
\ee
The function $U({x,\Im \tau_R})$ is given by:
\begin{align}\label{ufunction}
U(x,\Im \tau_R) &= {\re^{8x\log 2\over{ { 4 \pi} \Im \tau_R}}\over \sqrt{1+{ { 4 \pi} \Im \tau_R\over x}}} {Z_{\textrm{1-Loop}}\biggl(\sqrt{{x\over  { 4 \pi} \Im{\tau_R}}}\biggr)\over\sqrt{x\over { 4 \pi} \Im \tau_R}} { {\re}^{-12 \log \gamma_G+1+{1\over 3}\log 2}}\ec
\end{align}
and it has been constructed in such a way that $U(x,\Im \tau_R) $  is bounded and 
$U(x,\Im \tau_R) \to 1$   as $x\to\infty$. 
The reason for adopting such a rewriting of the matrix integral \eqref{detM} is to make contact with existing RMT literature. See appendix \ref{apcg} for relevant details.

In fact we could consider the family of matrix integrals labeled by two parameters $(\nu, \lambda)$:
\be\label{pertlag}
D^{(\nu, \lambda)}_{n} ={ 1\over n!} \int_{\bR^+} d^n x \prod_{i<j}(x_i-x_j)^2 \prod_i e^{-x_i}x_i^{\nu}(x_i+{4 \pi }\,{\Im \tau_R})^{\lambda}\ed
\ee
For $\nu=1$ and $\lambda=0$ we get: 
\be\label{approxd}
D^{(1,0)}_n = { 1\over n!} \int_{\bR^+} d^n x \prod_{i<j}(x_i-x_j)^2 \prod_i e^{-x_i}x_i\ec
\ee
which is the matrix model discussed at length in section \ref{largenpreds} obtained by truncating the expansion \eqref{largeR} at the first two contributions $\cF_0$ and $\cF_1$. 

At the end of section \ref{largenpreds} we hinted at the possibility that all the missing contributions in \eqref{approxd} comes from eigenvalues located near the origin. An efficient way to probe the effects  in the matrix model description is to add a ``singular perturbation'' by turning on a non-zero $\lambda$ as in \eqref{pertlag}. In this way the weight function has a modified behavior given by:
\be
x^\nu\left( x+{{ 4 \pi } \,{\Im \tau_R}}\right)^\lambda=
\begin{cases}
x^{\nu+\lambda} \ec &x\to \infty\ec\\
x^{\nu} \ec  &x\to 0\ed\\
\end{cases}
\ee
When $\nu=\lambda=\half$ we recover the structure appearing in \eqref{detM}. The matrix model is no longer exactly solvable but, as reviewed in appendix \ref{apcg}, there are interesting 
rigorous result regarding its large-$n$ asymptotics. In particular, the determinant \eqref{detM} has a large-$n$ limit which can be studied using the Coulomb gas approach. One finds (see appendix \ref{apcg}) :
\be\ba 
\det \cM_{(n)}  = \exp\biggl( &n^2\left(\log{n}-{3\over 2}-{ \log\left( { 4 \pi }\Im \tau_R \right)}\right)+n(- \log( { 4 \pi } \Im \tau_R ) -1 + \log{(2\pi)} + \log{n})
\\ &+{ n \left(12 \log \gamma_G-1-{1\over 3}\log 2\right)}+{7\over 48}\log{n} + \cO(1)\biggr)\ed\ea
\ee
Using the relation \eqref{presc} between $\det \cM_{(n)}$ and chiral ring data $G_{2n}(\tau,\bar \tau)$, we obtain the following large-$n$ asymptotics:
\be\ba 
\log G_{2n}(\tau, \bar \tau) =& 2n\log{n}  + 2\log{n} -2({ 1+\log\left({ 4 \pi}\Im \tau_R\right)})n +{\left(12 \log \gamma_G-1-{1\over 3}\log 2\right)} \\
&+  { \log{(2\pi)} -2 { \log\left({ 4 \pi} \Im \tau_R\right)}}+{47\over 48} {1\over n} + \cO\biggl({1\over n^2}\biggr)\ed
\ea\ee
The coefficient of the $1/n$ term now perfectly agrees with the EFT prediction~\eqref{expandedEFT}!
In addition, due to the strong constraints coming from the underlying Toda equation, this implies a  full agreement with the all-orders EFT prediction \eqref{EFTpred}.

We therefore have a lot of evidence now that the random matrix model in the limit of fixed $\Im\tau$ and large $n$ is dual to the non-relativistic Goldstone bosons theory. 
(The random matrix model is exact and also contains information about exponentially small corrections not visible in the EFT.) 
We now switch to studying the double scaling limit where $n/\Im\tau$ is kept fixed.

\section{A Double Scaling Limit}

From the viewpoint of ordinary perturbation theory, the limit of fixed $\Im\tau$ and large $n$ clearly exists. It is less obvious that the limit 
\be\label{doubles}
n, ~ \Im{\tau}  \to \infty\ec \qquad \lambda={n \over 4\pi \,{\Im}\tau}: \textrm{fixed}\ed 
\ee
exists. (Note that we redefined here $\lambda$ by a factor of $4\pi$ compared to previous notation in the paper.) This double scaling limit is reminiscent of the 't Hooft limit of gauge theories. Since we are studying $\SU(2)$ gauge theory, we do not have the analogs of double line notation and genus expansion and hence the existence of the limit~\eqref{doubles} is not obvious. In other words, it is not a priori guaranteed that the extremal correlators admit an expansion of the form 
\be\label{Nexpa}\log G_{2n} = \sum_{k=0}^\infty n^{1-k} C_k(\lambda)~. \ee 
Nevertheless, at small $\lambda$,  \cite{Bourget:2018obm, Beccaria:2018xxl,Beccaria:2018owt} have shown that perturbation theory re-organizes itself in a way consistent with~\eqref{Nexpa}. 
Our first goal is to give a conceptual proof for the existence of the expansion~\eqref{Nexpa}. 
Consider the random matrix theory with potential~\eqref{potential} rewritten in the double scaling limit~\eqref{doubles}
\be\label{threeterms}V={n\over \lambda} \Tr W -\frac12 \Tr \log W- \Tr \log Z_{\textrm{1-Loop}} (\sqrt  W)~. \ee 
The first term behaves exactly like a large $n$ theory of the 't Hooft type: we have a theory of $n\times n$ matrices with potential scaling like $n$~\cite{tHooft:1973alw}. 
The second and third terms have the notable property that they are single trace deformations that are down by one power of $n$ compared to the leading term. These two facts combined lead to a nice large $n$ limit because we can view the partition function as an expectation value $\langle e^{\cO }\rangle$ where $\cO$ is a single trace operator. In fact, to leading nontrivial order in the large $n$ expansion, 
\be\label{iden}\langle e^{\cO }\rangle = e^{\langle\cO\rangle } ~.\ee
This is because the largest contribution to a correlation function of single trace operators in the large $n$ limit is factorized (disconnected). 

If we keep just the first term in~\eqref{threeterms} we get an exactly solvable model of the type studied in previous sections. This model has a rather trivial dependence on $\lambda$ because the $\lambda$ dependence can be rescaled away. The first truly interesting contribution to the free energy is of order $n$ and arises from~\eqref{iden}, namely from $e^{\langle O\rangle}$. This leads to a contribution to $\log G_{2n}$ of order 1 in the double scaling expansion~\eqref{Nexpa} but that has a nontrivial $\lambda$ dependence, $C_1(\lambda)$. Our main goal would be to find $C_1(\lambda)$ and study its weak and strong coupling limits (we will also see that $C_0(\lambda)=2\log\lambda-2$).

\subsection{Matrix Model Analysis}\label{mmanalysis}

We need to organize the double scaling limit expansion for the matrix model with potential~\eqref{threeterms}. It is a little bit more convenient to regard the third term as the single trace interaction term and solve the model corresponding to the first two terms exactly. We therefore begin with
\be\label{threetermsi}V={n\over \lambda} \Tr W -\frac12 \Tr \log W~. \ee 
Note that this is not the same as the model considered in~\eqref{potentialsim} due to the factor of $1/2$ in front of the logarithm. But the model is solvable using the formulae given in that section. One would therefore like to first calculate 
\be D_n^{(1/2,0)}={1\over n!}\int_{\IR^n_+} \rd^n x \prod_{i=1}^n \re^{- {n\over \lambda}  x_i} \sqrt{x_i}\prod_{i<j}(x_i-x_j)^2~.\ee
This corresponds to computing the extremal correlators while omitting the $Z_{\textrm{1-Loop}}$ factor altogether.
Using~\eqref{normRMT} we find 
\be D_n^{(1/2,0)}={1\over n!}{\lambda^{n^2+n/2}\over n^{n^2+n/2}}\prod_{j=0}^{n-1} {\sqrt \pi \over 2^{2j+2} }\Gamma(2j+3)~.\ee
From this we can infer the contribution to the extremal correlators \footnote{We use the superscript  $\cN=4$  since this is essentially the problem we would have encountered in $\cN=4$ supersymmetric Yang-Mills theory with gauge group $\SU(2)$.}  $G_{2n}^{\cN=4}$  by following the usual prescription~\eqref{presc}. One has to be cautious because in taking the ratios of $D_n^{ (1/2,0)}$ we must keep $n/\lambda$ fixed and not $\lambda$. We therefore find that  
 \be G_{2n}^{ \cN=4} ={\sqrt \pi\over 2^{2n+2}(n+1)}\left({\lambda \over n}\right)^{2n+3/2}\Gamma(2n+3)~. \ee
 Taking the logarithm and expanding around large $n$ we find 
 \be\label{No1Loop} \log{G_{2n}^{\cN=4}}=  2n \left(\log(\lambda)-1\right)+{3\over 2}\log(\lambda) + \log(2\pi)  +{ \mathcal{O}(n^{-1})}\ed  \ee  
 We can therefore identify $C^{\cN=4}_0(\lambda)={2}\log(\lambda)-2\ec$ $C^{\cN=4}_1(\lambda)={3\over 2}\log(\lambda) + \log(2\pi)\ed$
 
As we argued above, the correction due to the third term in~\eqref{threeterms} must be subleading in $n$ and hence can at most contribute at the order $n^0$ for $G_{2n}$. Therefore, the result of~\eqref{No1Loop} that $C_0=2\log(\lambda)-2$ is exact. However, there are important corrections to $C_1$ which we will now analyze. 

In order to compute the corrections we need to evaluate the expectation value  
\be\label{correctionstoMSYM} \langle  Z_{\textrm{1-loop}}\rangle_{n} ={\int_{\IR_+} \rd^n x \prod_{i=1}^n \re^{- n  x_i} \sqrt{x_i}Z_{\textrm{1-loop}}\left(\sqrt{\lambda x_i}\right)\prod_{i<j}(x_i-x_j)^2\over \int_{\IR_+} \rd^n x \prod_{i=1}^n \re^{- n  x_i} \sqrt{x_i}\prod_{i<j}(x_i-x_j)^2}~.\ee
Since we are at present only interested in extracting $C_1(\lambda)$ we can use~\eqref{iden} to simplify the computation.
Note that, as explained in equation \eqref{iden},  in order to compute $\langle  Z_{\textrm{1-loop}}\rangle_{n}$ to leading nontrivial order we simply need to average $\log( Z_{\textrm{1-loop}})$ against the eigenvalue average density function. This density function already appeared before in~\eqref{MP} and it is known as the Mar\v{c}enko-Pastur density. We have to be careful because there we studied the large charge limit and here we are perturbing around~\eqref{threetermsi} which has a different coefficient in front of the logarithmic term. Consequently, The Mar\v{c}enko-Pastur density is slightly different:
\begin{equation}\label{MPi}
\begin{split}
\rho(x)&=\frac{\sqrt{(x-a) (b-x)}}{(2 \pi ) x}\ec\\
a &=-\frac{\sqrt{4 n+2}}{\sqrt{n}}+\frac{1}{2 n}+2\ec\\
b &= \frac{\sqrt{4 n+2}}{\sqrt{n}}+\frac{1}{2 n}+2\ed\\
\end{split}
\ee
We are ready to express the expectation value of $\log Z_{\textrm{1-loop}}$ as 
 \be  \langle  \log Z_{\textrm{1-loop}} \rangle_{n}=n \int_a^b \rd x \rho(x) \log Z_{\textrm{1-loop}}\left(\sqrt{x \lambda}\right)~, \ee
which is a nontrivial function of $n,\lambda$.
In order to derive the extremal correlator from $\langle  \log Z_{\textrm{1-loop}} \rangle_{n}$ we need to follow our usual prescription and take the ratio while remembering to vary $\lambda$.
{Let us first denote the normalized correlators as
\be\label{deltadef} \Delta {G}_{2n}\equiv {{G}_{2n}/ {G}_{2n}^{\cN=4}}\ed\ee }
 Hence,
\be\label{g2nex} \ba \log\Delta {G}_{2n}& ={\rd \over \rd n}\langle\log Z_{\textrm{1-loop}}\left(\sqrt{x \lambda}\right) \rangle_{n}\biggr|_{{\rm fixed}\  n/\lambda}+\mathcal{O}\left(n^{-1}\right)\\
&= \int_0^4 \rd x \rho_0(x) \left(\log Z_{\textrm{1-loop}}\left(\sqrt{x \lambda}\right)+\lambda \partial_\lambda \log Z_{\textrm{1-loop}}\left(\sqrt{x \lambda}\right) \right)+\mathcal{O}\left(n^{-1}\right)\ed\ea\ee
where 
\be  
\rho_0(x)={1\over 2\pi}\sqrt{\frac{4}{x}-1}\ed
\ee
Note that we have not used the more precise version of the eigenvalue density function~\eqref{MPi} because at the order of interest in the $1/n$ expansion, the edges of the distribution can be approximated by $0$ and $4$, respectively. 

Therefore, we now have a closed form formula for the contribution to $C_1(\lambda)$ from the interaction term (the third term) in the ensemble~\eqref{threeterms}: 
\be\label{genushalf}\Delta C_1(\lambda) =\int_0^4 \rd x \,\rho_0(x) \left(\log Z_{\textrm{1-loop}}\left(\sqrt{x \lambda}\right)+\lambda \partial_\lambda \log Z_{\textrm{1-loop}}\left(\sqrt{x \lambda}\right) \right)~. \ee
{Using \eqref{deltadef},} we can obtain the full answer for $C_1(\lambda)$ by combining $\Delta C_1(\lambda)$ with 
the contribution from $\cN=4$ SYM theory $C^{\cN=4}_1(\lambda)$ computed in~\eqref{No1Loop}:
\be  C_1(\lambda) =\Delta C_1(\lambda) +C^{\cN=4}_1(\lambda)~.\ \ee
 Therefore, we have a rather explicit representation of $C_1(\lambda)$. We can now study the weak coupling expansion and the strong coupling expansion\footnote{Interestingly, a conjecture for the all order weak coupling expansion (and resummation) of the quantity $C_1(\lambda)$ was already proposed in the paper \cite{Beccaria:2018owt}.} of~\eqref{genushalf}.

The weak $\lambda$ expansion is straightforward: 
\be\label{dz} \ba \Delta C_1(\lambda) 
&=\sum_{k\geq 2} \frac{i^{k+1} 2^{2-k} \left((-1)^k-1\right) \left(2^k-2\right) \pi ^{-k-\frac{3}{2}} \left(16 \pi^2 \lambda\right) ^{\frac{k+1}{2}} \zeta (k) \Gamma \left(\frac{k}{2}+1\right)}{(k+1)^2 \Gamma \left(\frac{k+1}{2}\right)}~.\
\ea\ee
Note that our general result for the weak coupling expansion~\eqref{dz} also explains the simplification in terms of the appearance of zeta numbers observed in \cite{Bourget:2018obm,Beccaria:2018xxl}. 

The analysis of the strong coupling limit is a little more technically challenging. Notice that, we can re-write\footnote{This expression can be derived, for example, starting from the series expansion at weak coupling \eqref{dz}. In doing so, we found particularly useful to consult chapters 3 and 4 of the book \cite{paris_kaminski_2001}. This was inspired by~\cite{Binder:2019jwn}.} the integral~\eqref{genushalf} as 
\be\label{c1int} \Delta C_1(\lambda) = \int_0^{\infty}{\rd t}\,\frac{4 \left(-4 \lambda  t^2+4 \sqrt{\lambda } t J_1\left(4 t \sqrt{\lambda }\right)+J_0\left(4 t \sqrt{\lambda }\right)-1\right)}{\left(e^t+1\right) t^2}\ec \ee
with $J_0(x)$ and $J_1(x)$ ordinary Bessel functions. This rewriting facilitates the study of the strong coupling limit. Experimentally we find that the integral \eqref{c1int} can be written as:
\be\ba\label{g2nstrong}   \Delta C_1(\lambda)&=12 \log ({ \gamma_{G}})-1-\frac{\log (2)}{3} -16 \lambda \log (2)+{1\over 2}\log \lambda + F^{\rm inst}(\lambda)~,\ea\ee
where
\be \label{guessinst}\ba   F^{\rm inst}(\lambda)&= \sum_{n\geq0 }\frac{8 \left(K_0\left(4 (2 n+1) \pi  \sqrt{\lambda }\right)+4 \pi  \sqrt{\lambda } (2 n+1) K_1\left(4 (2 n+1) \pi  \sqrt{\lambda }\right)\right)}{(2 \pi  n+\pi )^2}\ec \\
& = \re^{-4 \pi \sqrt{\lambda}}\left(\frac{11}{2 \sqrt{2} \pi ^2 \sqrt[4]{\lambda }}+\frac{8 \sqrt{2} \sqrt[4]{\lambda }}{\pi }-\frac{31}{128 \sqrt{2} \pi ^3 \lambda ^{3/4}}+\mathcal{O}\left(\lambda^{-5/4}\right)\right) +\mathcal{O}\left(\re^{-12 \pi \sqrt{\lambda}}\right)\ea\ee
where $K_0(x)$ and $K_1(x)$ are modified Bessel function of the second kind. 


Let us now discuss the physical meaning of the strong coupling result. 
As explained in the introduction, the physics of the double scaling limit is that of the BPS particles with mass of order $\sim \sqrt{\lambda}$. This is why we get a nontrivial dependence on $\lambda$, in contrast to the standard large charge limit. When we take $\lambda\to\infty$ (but much smaller than $n$) we are making the BPS particles heavy. As a result, the exponentially small term 
\be\label{insta} \frac{8 \sqrt{2} \sqrt[4]{\lambda }}{\pi }\re^{-4 \pi \sqrt{\lambda}}\ec\ee 
could perhaps be interpreted as the worldline of the hypermultiplet BPS particle.\footnote{The exponent in~\eqref{insta} is indeed consistent with the propagation of a massive hypermultiplet. We would like to thank S. Hellerman for several conversations on the topic.} From this point of view it is puzzling that \eqref{g2nstrong} does not have a contribution that scales like $\re^{-8 \pi \sqrt{\lambda}}\,$: this is perhaps due to a cancelation between the disconnected two-hyper contribution and the $W$ boson.\footnote{The piece~${1\over 2}\log \lambda$ in \eqref{g2nstrong} is easy to understand -- it combines with  ${3\over 2}\log \lambda$ in~\eqref{No1Loop} to give $2\log\lambda$, which is the correct coefficient to reproduce the $a$-anomaly on the Coulomb branch~\eqref{almostthere}.}

One very interesting question that we do not address here is the connection to non-perturbative terms in the fixed $\Im\tau$ limit. At least for small enough $g_{\rm YM}$,  the limit of fixed coupling and large charge is quite similar to the limit of large $\lambda$. However, in the large charge limit with  fixed $\Im\tau$ we could also have non-perturbative effects with scaling
\be  \re^{-A \sqrt{{n \Im \tau}}} \ec \ee
corresponding to the propagation of monopoles and thus invisible in the 't Hooft expansion. An analogous example of such phenomenon can be found, for example, in ABJM theory. In that context there are two distinct sources of non-perturbative effects which can be interpreted as worldsheet instantons (suppression in $\lambda$) and membrane instantons (suppression in $g_s$) \cite{Drukker:2010nc, Drukker:2011zy, Grassi:2014vwa}. 

Another question which we discuss only very briefly is the omission of the instanton terms in~\eqref{pestunper}. In the limit with fixed coupling and large $n$ it is hard to see why a priori this truncation is justified. Yet, we have argued that to all orders in the $1/n$ expansion of $G_{2n}$ the results agree with the predictions from effective field theory. However, it seems rather unlikely that the non-perturbative terms in this limit would be correctly captured by the truncated partition function~\eqref{pestunper}.  The situation in the double scaling limit is under much better control. 
The original gauge instantons are suppressed by the instanton fugacity $q$ which in the double scaling limit translates to $e^{-2\pi n/\lambda}$ with $\lambda$ fixed in the double scaling limit as usual. Therefore, the omission of $\SU(2)$ instantons does not affect the double scaling expansion~\eqref{Nexpa} to any order and the truncation~\eqref{pestunper} is therefore very well justified mathematically. In particular, we can really trust the non-perturbative terms in the 't Hooft coupling such as~\eqref{insta}.

\section{Random Matrix Theory Approach to Argyres-Douglas Theory}\label{RMTAD}
\subsection{Elements of Seiberg-Witten Theory}\label{SWelements}
The Seiberg-Witten \cite{Seiberg:1994rs, Seiberg:1994aj} IR solution for $\cN=2$ theories with a rank 1 Coulomb branch, parametrized by a complex coordinate $u \in \bC$, is described in terms of 
an Abelian $\cN=2$ vector multiplet. At leading order all interactions are governed by a special K\"ahler sigma model with a holomorphic prepotential $\cF$ which is a function of the vector multiplet scalar component $a$. In order study the low energy dynamics, it is very convenient to introduce a new variable $a_D \equiv {\partial \cF(a) \over \partial a}$. The metric on the moduli space of vacua can be expressed in terms of a set of special coordinates denoted by $(a, a_D)$. More precisely, $a$ and $a_D$ are non-trivial functions of $u$ which can be obtained by performing some definite integrals also known as period integrals. 


The dependence on $u$ can be inferred from the explicit expressions of $(a(u), a_D(u))$ obtained from the period integrals. For example, in the $\cN=2$ theory with $N_f=4$ massless hypers there is a simple relation given by
\be\label{matoneNf4}
u = a^2+\mathcal{O}(q)\ec
\ee
where $q$ is the instanton fugacity defined in section \ref{gaugeNf4}. The above analysis can be generalized to other rank 1 $\cN=2$ theories and this will be particularly important for extending our study of extremal correlators since taking a $\tau$ and $\bar \tau$ derivative as in \eqref{matrixdef} can be interpreted in terms of an insertion of $u$ inside the $S_4$ partition function. Here, we will focus on the simplest class of Argyres-Douglas  superconformal fixed points arising as special points on the moduli space of $\SU(2)$ gauge theories with $N_f \leq 3$ massive fundamental flavors \cite{Argyres:1995jj, Argyres:1995xn}. We will now review some basic aspects of this construction with focus on the $N_f=1$ theory.

Let us begin from a $\cN=2$ theory with a single fundamental massive fundamental hypermultiplet. The IR curve for this theory has been written a long time ago in \cite{ Seiberg:1994aj}:
\be 
\label{sw}y^2=x^4-\Lambda_1^3 m + u^2 - \Lambda_1^3 x - 2 u x^2 \ec
\ee
where $\Lambda_1$ is the UV scale, $m$ is the hypermultiplet mass and $u$ parametrize the Coulomb branch as before. For $m=m_* \equiv {3\over4} \Lambda_1$ the discriminant of the curve \eqref{sw} has a double zero at:
\be
u=u_{*}\equiv {3\over 4}\Lambda^2_1\ed
\ee
When both $m=m_*$ and $u=u_*$, mutually non-local particles become massless and the low-energy theory cannot have a Lagrangian description. 
Near the singularity, the theory can be modeled by an IR curve which reads \cite{Argyres:1995xn}:
\be \label{defad} 
y^2=(x^3 -3 \Lambda_{\rm AD}^2 x + u_{\rm AD})\ed
\ee 
The superconfomal fixed point is reached by tuning $\Lambda_{\rm AD}\to 0$ at the origin of the Coulomb branch $u_{\rm AD}=0$. From the curve \eqref{defad} it is possible to determine the $R$-charge scaling of $u_{\rm AD}$ which, in this case, is ${6\over 5}$. The corresponding generator of the Coulomb branch chiral ring has $\Delta_{\cO}={6\over 5}$. 
The analysis of period integrals for the $AD_{N_f=1}(\SU(2))$ theory can be done as in \cite{Masuda:1996xj,Masuda:1996np} where we first set $m=m_*$ and {then analytically continue the expression for $(a(u), a_D(u))\bigr|_{m=m_*}$ near a region  $u=u_*$}:
\be
\begin{split} \label{aad}
\overline{a}(u) &={ a_*}+ {{3 \sqrt{3} \ri}  \Lambda_1 \over {8} \sqrt{  y}} \Biggl(\frac{6\, (-y)^{-2/3} \Gamma \left(-\frac{1}{3}\right) \, _3F_2\left(\frac{2}{3},\frac{2}{3},\frac{7}{6};\frac{4}{3},\frac{13}{6};\frac{1}{y}\right)}{7 \Gamma \left(\frac{1}{3}\right)^2}\\ &\hspace{6cm}+ \frac{{6 (-y)^{-1/3}} \Gamma \left(\frac{1}{3}\right) \, _3F_2\left(\frac{1}{3},\frac{1}{3},\frac{5}{6};\frac{2}{3},\frac{11}{6};\frac{1}{y}\right)}{5 \Gamma \left(\frac{2}{3}\right)^2}\Biggr)\ec\\
\overline{a}_D(u) &=-{{9 \ri \pi}  \Lambda_1 \over {8} \sqrt{y}} \Biggl(\frac{{6 (-y)^{-1/3}} \Gamma \left(\frac{1}{3}\right) \, _3F_2\left(\frac{1}{3},\frac{1}{3},\frac{5}{6};\frac{2}{3},\frac{11}{6};\frac{1}{y}\right)}{5 \Gamma \left(\frac{2}{3}\right)^2}\ec\\
&\hspace{6cm}- {6 (-y)^{-2/3}\Gamma \left(-\frac{1}{3}\right)\, _3F_2 \left(\frac{2}{3},\frac{2}{3},\frac{7}{6};\frac{4}{3},\frac{13}{6};\frac{1}{y}\right)\over 7 \Gamma \left(\frac{1}{3}\right)^2}\Biggr)\ec
\end{split}
\ee
where\footnote{The overline notation used here has been chosen to emphasize the analytic continuation.} 
\be y \equiv {27 \over 4}\left(3-{4 u  \Lambda_1^{-2} }\right)^{-1}\ec\qquad a_* \equiv {3\Lambda_1 \over 4}\ed \ee
with $a_*$ denoting the value of $a$ at the AD point. The dependence on $u$ in \eqref{aad} can be now inverted to find a generalization of \eqref{matoneNf4} for the  $AD_{N_f=1}(\SU(2))$ theory. By expanding \eqref{aad} around  $u=u_*$ we find
\be \label{aud} 
\overline{a}(u_*+\Lambda_1 ^2 u_{\rm AD})-a_* = \Lambda_1 \frac{2^{4/3} \Gamma \left(\frac{1}{3}\right)}{5 \Gamma \left(\frac{2}{3}\right)^2} u_{\rm AD}^{5/6}+ \cO(u_{\rm AD}^{7/5})\ed
\ee

\subsection{Extremal Correlators in Argyres-Douglas Theories}\label{extad1}

In this section we present the calculation of large charge extremal correlators using random matrix model techniques. The scaling limit to the AD point that we reviewed above will be a key element in the discussion. Let us  start from the UV supersymmetric $S^4$ partition function for a $\SU(2)$ $\CN=2$ gauge theory with $N_f=1$ 
\be \label{s4pf}
Z_{S^4}[\tau, \bar \tau]=\int _{{  \bR}}\rd a{ (2a )^2 H(2 \ri a R)^2\over H(\ri(a+m)R)H( \ri(-a+m)R)}\re^{-4 \pi {\rm Im}\tau R^2a^2}|Z_{\rm Inst}(  \ri a, \Lambda_1, m,R)|^2\ec
\ee
where $m$ denotes the hypermultiplet mass. The instanton partition function is now given by 
\be \label{zi}Z_{\rm Inst}( \ri a, \Lambda_1, m,R)=1+\frac{\Lambda_1^3 m R^4}{8 a^2 R^2+8}+\frac{\Lambda_1^6 R^6 \left(R^2 \left(a^2 \left(8 m^2 R^2-3\right)+33 m^2\right)-3\right)}{64 \left(a^2 R^2+1\right) \left(4 a^2 R^2+9\right)^2}+\mathcal{O}(\Lambda_1^9)\ed\ee

Following our analysis from section \ref{largeradius} we can now present the large charge expansion.\footnote{Our analysis in also inspired by \cite{Russo:2014nka} where it was found that, at large radius, the AD fixed point is a saddle of the $S^4$ partition function for the $\SU(2)$, $N_f=2$ theory.} The idea, as before, is to only retain contributions from $\cF_0$ and  $\cF_1$.\footnote{Here  ${\cF}_0$ contains the full quadratic factor appearing in the classical contribution to the $S^4$ partition function. In addition we set $\exp(-4 \pi\, {\Im}\tau)= { e^{6+6 \gamma } \over 16}R^6 \Lambda_1 ^6. $}  At $m=m_*$ we can write
\be \label{f0f1cf}
\begin{split}
 \partial_a {\cF}_0 & = \overline{a}_D(u)\ec \\
 \cF_1& =  \frac{1}{12} \left(\log \left( \frac{\left(3 \Lambda_1^2-4 u\right)^2 \left(15 \Lambda_1^2+16 u\right)}{2^{16}}\right)-6 \log \left( \left(\overline{a}(u)\right)^2 {\rd \overline{a}(u)\over \rd u}\right)\right)\ed\\
\end{split}\ee
We would like to stress that in order to describe the Argyres-Douglas superconformal fixed point we need to obtain a closed analytic form for $\cF_0$ and $\cF_1$
by resumming the instanton series {as in \eqref{f0f1cf}}. (This is to be contrasted with $\SU(2)$ $N_f=4$ superconformal theory, where one could take a weak coupling limit.) In the language of section \ref{SWelements} we find that, for $m=m_*$\footnote{In the topologically twisted partition function for $AD_{N_f=1}(\SU(2))$ \cite{Moore:2017cmm} one should also consider the effects of perturbing the mass away from $m=m_*$. 
In the large charge expansion however these effects can be neglected.} in a neighborhood of $u_*$, $\cF_0$ and $\cF_1$ have the following scaling behavior
\be
\begin{split}
 \partial_a {\cF}_0(u_*+\Lambda_1^2u_{\rm AD})&= - \frac{2^{4/3} \sqrt{3} \pi  \Gamma \left(\frac{1}{3}\right)\Lambda_1}{5 \Gamma \left(\frac{2}{3}\right)^2} \,u_{\rm AD}^{5/6}+ \cO(u_{\rm AD}^{7/5})\ec\\
a^2 \exp\left(2 \,\cF_1\right) (u_*+\Lambda_1^2u_{\rm AD})&=  \frac{\sqrt{3} \pi ^2 \Lambda_1^2}{\sqrt[3]{2} \Gamma \left(\frac{1}{3}\right)^3}\,u_{\rm AD}^{1/2}+\mathcal{O}(u_{\rm AD}^{\,{5 \over 6}})\ed
\end{split}
\ee
Hence, by \eqref{aud}, we find that the infinite-dimensional matrix $\cM_{k, l}$ generating the extremal correlators can be approximated as follows (keeping only $\cF_0$ and $\cF_1$) for the Argyres-Douglas fixed point:
\be\label{myg2nb} 
\cM_{k,l}= 
 \mathbf{N_\cO}(\Lambda_1, R)\int_{\bR} da\left(|a|^{6/5}\right)^{k+l}a^{3/5} \exp\left(- a^2 \right) \ec
 \ee
where $\mathbf{N_\cO}(\Lambda, R)$ is a normalization factor whose explicit form will not be important for our purposes. 
We can now repeat the same steps which we described in subsection \ref{sqcdrm} and obtain an emergent random matrix model description for the $AD_{N_f=1}(\SU(2))$ theory:
\be\label{matAD1}
\det{\cM_{(n)}}={ 1\over n!} \int_{\bR^+} \rd^n x \prod_{i<j}(x_i-x_j)^2 \prod_i e^{-x_i^{5/3}}x_i^{1/3}\ed
\ee
As far as we know, the above matrix model does not have an analytic solution. But we are still able to show that the leading large $n$ behavior for extremal correlators is governed by:
\be\label{largeAD1}
\log G_{2n} = {6 \over 5}\,n\log n + {4\over 5} \log n + \cO\left({1\over n}\right)\ec
\ee
where all the non-universal terms depending on normalizations have been dropped. The above formula matches perfectly with effective field theory predictions obtained by plugging the values for $AD_{N_f=1}(\SU(2))$, listed in table \ref{tab:alphavalues}, in equation \eqref{EFTpred}.  This therefore gives direct evidence that the large charge limit of BPS states in $AD_{N_f=1}(\SU(2))$ is described by an effective theory for the Coulomb branch field.

We hope to report on the computation of higher-order terms in~\eqref{largeAD1} elsewhere.

\subsubsection{Extension to $AD_{N_f=2}(\SU(2))$ and $AD_{N_f=3}(\SU(2))$ Fixed Points}
The ideas presented in section \ref{SWelements} and \ref{extad1} can be readily generalized to other rank 1 Argyres-Douglas superconformal fixed points which arise as special points on the Coulomb branch of $\cN=2$ $\SU(2)$ theories with $N_f=2, 3$ massive hypermultiplets. For all these rank 1 models, the general form of the matrix model is given by 
\be\label{AD23}
{\det{\cM_{(n)}}}={ 1\over n!} \int_{\bR^+} \rd^n x \prod_{i<j}(x_i-x_j)^2 \prod_i e^{-x_i^{\upalpha}}x_i^{\upbeta}\ec
\ee
 where\footnote{The relevant values of $\Delta_\cO$ and $\alpha$ for these theories can be found in table \ref{tab:alphavalues}.} 
\be
\upalpha \equiv {2\over \Delta_{\cO}}\ec \qquad \upbeta\equiv {1+ 2\alpha\over\Delta_{\cO}} -1\ed
\ee
As before, we can calculate the large charge behavior of \eqref{AD23} and find perfect agreement with effective field theory up to $\cO\left({1\over n}\right)$. 

A natural question which we might pose now is whether the large $n$ results of the matrix model extrapolate reasonably well to lower values of $n$. 

\subsection{Comparison with Numerical Bootstrap}
\renewcommand{\arraystretch}{1.6}
\renewcommand\tabcolsep{6pt}
\begin{table}[t]
  \centering
  \begin{tabular}{ |c|c| }
\hline
{\bf Theory} &  {$\lambda^2_{\cO}$}  \\
\hline
\hline
$AD_{N_f=1}(\SU(2))$&  $2.0982$ \\
\hline
$AD_{N_f=2}(\SU(2))$&  $2.2412$  \\
\hline
$AD_{N_f=3}(\SU(2))$&  $2.4206$ \\
\hline
\end{tabular}
  \caption{Numerical values of the OPE coefficient $\lambda^2_{\cO}$ obtained from the RMT approach (including only $ \cF_0 $ and $\cF_1$).}
  \label{tab:OPE}
\end{table}
It was shown in \cite{Cornagliotto:2017snu} that the simplest Argyres-Douglas superconformal fixed point, $AD_{N_f=1}(\SU(2))$, has an OPE coefficient $\lambda^2_{\mathcal{E}_{12/5}}$ which satisfies the following rigorous bound:
\be 
\label{1711b}2.1418 \leq \lambda^2_{\mathcal{E}_{12/5}} \leq 2.1672 \ed
\ee
The above coefficient can be calculated with the matrix model approach from section \ref{extad1} using the following combination of extremal correlators:
\be\label{OPEnum}
\lambda^2_{\cO} = \frac{G_4G_0}{G_2^2}\ec
\ee
which is completely independent of normalisation factors.\footnote{The OPE coefficient  \eqref{OPEnum} can be computed analytically since the computation only involves taking determinants of  $4\times 4$ and $2\times 2$ matrices or, equivalently, solving the matrix model integral for $2$ or $1$ integration variables. 
} The matrix model approach gives a uniform description for all rank 1 Argyres-Douglas fixed points, in particular it can be used to calculate the value of the OPE coefficient also in $AD_{N_f=2}(\SU(2))$ and $AD_{N_f=3}(\SU(2))$ fixed points which do not yet have a tight numerical bootstrap bound.

 We collect all the numerical values of OPE coefficients \eqref{OPEnum} in table \ref{tab:OPE}. 
The predicted value of $\lambda^2_{\cO}$ in $AD_{N_f=1}(\SU(2))$ calculated with the matrix model \eqref{matAD1} is below the lower end of the bootstrap bound approximately just by $2\%$!

We would like to make two comments about this discrepancy:
First, since our approximation scheme for the matrix model is based solely on $\cF_0$ and $\cF_1$, it should not come as a surprise that the matrix model does not work perfectly for small $n$, if anything, it is surprising that the discrepancy is so small. 
Second, from the spectrum of BPS particles on the Coulomb branch one could estimate the effects of worldline instantons and try to fit the small $n$ values better by including some non-perturbative corrections.

\bigskip

\begin{center}
\large\textbf{Acknowledgments}
\end{center}
\vspace{-3pt}

\noindent We are grateful to A. Abanov, M. Beccaria, C. Cordova, M. Del Zotto, A. Dymarsky, J. Gu, S. Hellerman, E. Kanzieper, M. Lemos, D. Mazac, A. Neitzke, N. Nekrasov, D. Orlando, N. Piazzalunga, V. Saxena, M. Serone,  and S. Shatashvili,  for discussions. 
Z.K. is supported in part by the Simons Foundation grant 488657 (Simons Collaboration on the Non-Perturbative Bootstrap). Any opinions, findings, and conclusions or recommendations expressed in this material are those of the authors and do not necessarily reflect the views of the funding agencies.

\bigskip

\appendix

\section{Coulomb Gas for Perturbed Laguerre Ensemble }\label{apcg}

A central tool in the analysis of section \ref{largenpreds} is the following matrix integral
\be \label{d0}D^{(\nu, \lambda)}_{n} = {1\over n!}\int_{\bR_+^n} d^n x \prod_{i=1}^n e^{-x_i}x_i^{\nu}(x_i+t)^{\lambda}  \prod_{i<j}(x_i-x_j)^2\ed \ee
We would like to analyze the large $n$ asymptotic of \eqref{d0} in the Coulomb gas formalism following the works  \cite{Chen_1998, hank,hank2, Chen1}. A pedagogical review of the 
Coulomb gas, geared towards RMT, can be found in \cite{2017arXiv171207903L}. Let us write \eqref{d0} as
\be\label{dddd1} D^{(\nu, \lambda)}_{n} = n ^{n\lambda } Z^{(\nu, 0)}_{n} {Z^{(\nu, \lambda)}_{n} (t)\over Z^{(\nu, 0)}_{n} }\ec\ee
where 
\be \label{zi}Z^{(\nu, \lambda)}_{n} (t)= {1\over n!}\int_{\bR_+^n} d^n x \prod_{i<j}(x_i-x_j)^2 \prod_{i=1}^n e^{-n x_i}x_i^{\nu}(x_i+t n^{-1})^{\lambda}.\ee
{Note that $Z^{(\nu, 0)}_{n} $ can be evaluated exactly by using \eqref{cwl}. It is however instructive to } consider the large $n$ limit of $Z^{(\nu, 0)}_{n}$. The equilibrium density for the eigenvalues $\sigma(x)$ {is, as anticipated in the main text, the Mar\v{c}enko-Pastur law }
\be \sigma(x)={1\over 2 \pi x} \sqrt{(b-x)(x-a)}\ec \qquad a\leq x\leq b \ec\ee
where $(a,b)$ are also known as ``endpoints of the cut." These can be evaluated by standard matrix model techniques
\be 
\begin{split} 
b&= \frac{\nu +2 \sqrt{n (\nu +n)}+2 n}{n}=4+\mathcal{O}\left(\frac{1}{n}\right)\ec\\
 a&=\frac{\nu -2 \sqrt{n (\nu +n)}+2 n}{n}=\frac{\nu ^2}{4 n^2}+\mathcal{O}\left({1\over n^{3}}\right) \ed
 \end{split}
 \ee
It is also convenient to think about the quantity appearing on the right hand side of \eqref{dddd1} as an expectation value:
\be  {Z^{(\nu, \lambda)}_{n} (t)\over Z^{(\nu, 0)}_{n} } = \biggl\langle \left(x+{t\over n}\right)^{\lambda\,}\biggr\rangle_{(\nu, 0)}\ed\ee
At large $n$ the above expectation value is approximated by  \cite{Chen_1998, hank,hank2, Chen1}:
\be  {Z^{(\nu, \lambda)}_{n} (t)\over Z^{(\nu, 0)}_{n} }\approx \exp\left[-\lambda^2{S_1(n,t)\over 2}-\lambda S_2(n,t)\right]\ec \ee
where
  \be S_1(n,t)=-2 \log \left(\frac{1}{2} \Biggl(\frac{b+t n^{-1}}{a+t n^{-1}}\Biggr)^{1\over 4}+\frac{1}{2} \Biggr( \frac{a+t n^{-1}}{b+t n^{-1}}\Biggl)^{1\over 4}\right)\ec  \ee
and
\be
\begin{split}
S_2(n,t)&= \frac{1}{2} n  \sqrt{a b} \log \left(\frac{\left(\sqrt{(a+t n^{-1}) (b+t n^{-1})}+\sqrt{a b}\right)^2-t^2 n^{-2}}{\left(\sqrt{a}+\sqrt{b}\right)^2}\right)\\
&+\frac{1}{4} n \left(-2 \sqrt{a+t n^{-1}} \sqrt{b+t n^{-1}}+a+b+2 t n^{-1}\right) \\
&-\frac{n}{2}  (a+b) \log \left(\frac{1}{2} \left(\sqrt{a+t n^{-1}}+\sqrt{b+t n^{-1}}\right)\right)\ed\end{split} 
\ee
It follows that, taking the large $n$ limit while keeping $t$ fixed, \eqref{dddd1} is given by
\be
\begin{split}
\label{dnl} D^{(\nu, \lambda)}_{n} &=\exp\left[n^2 \log n -{3/2} n^2 +{n}(\nu+\lambda)\log n + (-\nu-\lambda+\log(2\pi)) n +\right. \\
\qquad &+ \left. 2 \lambda  \sqrt{n t}+ \left({\nu^2\over 2}-{1\over 6}+{\lambda^2\over 4}+{\lambda \nu \over 2}\right)\log n +\mathcal{O}(1)\right] \ed
\end{split}
\ee
We can now substitute $(\nu=\half, \lambda = \half)$ into \eqref{dnl} and get
\be\begin{split}\label{ddd}D^{(1/2, 1/2)}_{n} =\exp\left[n^2 \log n -{3/2} n^2 + n\log n + ({-1}+\log(2\pi)) n + \sqrt{n t}+ {7\over 48}\log n +\mathcal{O}(1)\right]\ed\end{split} \ee
The large $n$ analysis presented above is particularly useful for the study of $\cN=2$ $\SU(2)$ gauge theory with  $N_f=4$ at large charge. The basic idea is that the matrix integral
\eqref{detM}  can be seen as a perturbation of   $D^{(1/2, 1/2)}_{n} $ by a function 
 $U$ which is bounded on the positive axis and goes to 1 sufficiently fast as $x\to \infty$.
We are interested in calculating how such perturbation affects the large $n$ asymptotic of \eqref{ddd}.  Interestingly this type of questions were successfully addressed in the RMT literature, see for instance \cite{hank,hank2,Chen1}. In these works, the unperturbed ensemble is taken to be $D^{(\nu, 0)}_{n} $. However, their analysis carry over to the case of $D^{(\nu, \lambda)}_{n} $ as well.
After some lengthy algebra we find\footnote{
Strictly speaking the  $\mathcal{O}(1)$ in \eqref{tu}  was proven mathematically in \cite{hank} for functions $U(x)$ such that $U(x)-1$ is a Schwartz function. In \cite{hank,hank2} however it is argued (in a slightly less rigorous way) that the Schwartz condition can be relaxed to include functions $U(x)$ which are bounded, sufficiently regular and approaching the value 1 at infinity.
} 
\be  \label{tu}\begin{split} \left\langle U \right\rangle_{(1/2,1/2)} & = {\int_{\bR^+} d^n x \prod_{i<j}(x_i-x_j)^2 \prod_{i=1}^n e^{-n x_i}x_i^{1/2}(x_i+{4 \pi}n^{-1}\,{\Im \tau_R })^{1/2}\,U(x_i n, { \Im \tau_R})\over {\int_{\bR^+} d^n x \prod_{i<j}(x_i-x_j)^2 \prod_{i=1}^n e^{-n x_i}x_i^{1/2}(x_i+{4 \pi}n^{-1}\,{\Im \tau_R})^{1/2} }}\\
& = \exp\left[-\sqrt{4 \pi \Im \tau_R} \sqrt{n}+ \mathcal{O}(1)\right] ,\end{split}\ee
In deriving this expression, we had to use an interesting identity 
\be \label{id1l}\int_0^{\infty}\log\left(\re^{8x^2\log 2} x^{-1} {Z_{\textrm{1-Loop}}({{x}})} { {\re}^{-12 \log \gamma_G+1+{1\over 3}\log 2}} \right) {\rd x}=0\ed\ee
Finally, we  write \eqref{detM}  as
\be \det \cM_{(n)} = { {{\re}^{ 12 n \log \gamma_G-n-{n \over 3}\log 2}}\over{\left(4 \pi \Im \tau_R\right)^{n(n+1)}}} D^{(1/2, 1/2)}_{n} 
\left\langle U \right\rangle_{1/2,1/2}\ec\ee
where  in $D^{(1/2, 1/2)}_n$ we set $ t= 4 \pi \Im \tau_R$ .
Hence from \eqref{tu} and \eqref{ddd} we find that
\be \ba \label{masy}  \det \cM_{(n)}  =\exp &\left[  n^2 \log n -{3/2} n^2 + n\log n + ({ -2}+\log(2\pi)+12  \log \gamma_G-{1 \over 3}\log 2) n \right.\\
&\left. + n(n+1) \log \right(4 \pi \Im \tau_R\left)+ {7\over 48}\log n +\mathcal{O}(1)\right]\ed\ea \ee
The identity \eqref{id1l} is crucial for having only integer powers of $n$ in the asymptotics \eqref{masy}. (If not for this mysterious identity, effective field theory would not be correct.)

\newpage
%
\bigskip
\renewcommand\refname{\bfseries\large\centering References\\ \vspace{-0.4cm}
\addcontentsline{toc}{section}{References}}
\bibliographystyle{utphys.bst}
\bibliography{CoulombBranchFinalV1Fixed.bib}
	
\end{document}